\documentclass[preprint,
superscriptaddress,
 amsmath,amssymb, aps, prx
]{revtex4-2}

\usepackage[T1]{fontenc}
\usepackage[utf8]{inputenc}
\usepackage{lmodern}
\usepackage{amsmath,amssymb,mathtools,bm}
\usepackage{booktabs}
\usepackage{array}
\usepackage{graphicx}
\usepackage{xcolor}
\usepackage{braket}
\usepackage[section]{placeins}

\usepackage{threeparttable}

\newcommand{\threesubsection}[1]{%
  \medskip
  \textit{#1}: }

\newcounter{algorithm}
\newenvironment{algorithm}[1][]{%
  \par\addvspace{\medskipamount}%
  \refstepcounter{algorithm}%
  \renewcommand{\caption}[1]{%
    \par\noindent\textbf{Algorithm~\thealgorithm:} ##1\par\nobreak\smallskip}%
}{%
  \par\addvspace{\medskipamount}%
}
\newcounter{alglinecnt}
\newenvironment{algolines}{%
  \begin{list}{\footnotesize\arabic{alglinecnt}:}{\usecounter{alglinecnt}%
    \setlength{\itemsep}{1pt}\setlength{\parsep}{0pt}%
    \setlength{\labelsep}{0.8em}\setlength{\leftmargin}{2.4em}}\small}%
  {\end{list}}
\newcommand{\algI}{\hspace*{1.6em}}   
\newcommand{\algII}{\hspace*{3.2em}}  

\usepackage{hyperref}
\usepackage{cleveref}
\crefname{figure}{Fig.}{Figs.}
\Crefname{figure}{Figure}{Figures}
\crefname{equation}{Eq.}{Eqs.}
\Crefname{equation}{Equation}{Equations}
\crefname{section}{Sec.}{Secs.}
\Crefname{section}{Section}{Sections}
\crefname{subsection}{Sec.}{Secs.}
\Crefname{subsection}{Section}{Sections}
\crefname{table}{Tab.}{Tabs.}
\crefname{algorithm}{Algorithm}{Algorithms}
\Crefname{algorithm}{Algorithm}{Algorithms}
\Crefname{table}{Table}{Tables}

\begin{document}

\title{Probing the Weak-Driving Quantum Speed Limit via Drift-Aware Shooting Methods}

\author{Denis Jankovi\'c}
\email{denis.jankovic@qns.science}
\affiliation{Center for Quantum Nanoscience, Institute for Basic Science, Seoul 03760, Republic of Korea}
\affiliation{Ewha Womans University, Seoul 03760, Republic of Korea}

\author{Saba Taherpour}
\affiliation{Center for Quantum Nanoscience, Institute for Basic Science, Seoul 03760, Republic of Korea}
\affiliation{Department of Physics, Ewha Womans University, Seoul 03760, Republic of Korea}

\author{Paul-Louis Etienney}
\affiliation{Universit\'e de Strasbourg, CNRS, Institut de Physique et Chimie des Mat\'eriaux de Strasbourg, UMR 7504, 67000 Strasbourg, France}

\author{Paul-Antoine Hervieux}
\affiliation{Universit\'e de Strasbourg, CNRS, Institut de Physique et Chimie des Mat\'eriaux de Strasbourg, UMR 7504, 67000 Strasbourg, France}

\author{Christoph Wolf}
\email{wolf.christoph@qns.science}
\affiliation{Center for Quantum Nanoscience, Institute for Basic Science, Seoul 03760, Republic of Korea}
\affiliation{Ewha Womans University, Seoul 03760, Republic of Korea}

\begin{abstract}
A central goal of quantum optimal control is to achieve high-fidelity and low-energy control pulses. When quantum optimal control methods optimize every point of a pulse discretized over small time steps independently this can yield high fidelity control but also results in broadband and energy-hungry waveforms. We extend MAGICARP --- a shooting method inspired by Pontryagin's maximum principle on energy that generates an entire pulse from a small set of parameters, making it smooth and energy-efficient by construction --- from driftless systems to closed systems with the constant drift Hamiltonian of two exchange-coupled spins in an external magnetic field. The optimization proceeds in stages: the dressed states of the drift Hamiltonian structure the target, an initial shooting optimization is performed in the rotating-wave frame, and an exact laboratory-frame refinement follows. Benchmarked against Krotov and GRAPE at matched gate infidelity, MAGICARP consistently achieves the lowest energy and a conserved pulse area, concentrates its spectral weight on the gate-relevant transitions, and is the most robust to fluctuations in the exchange coupling; GRAPE independently converges to essentially the same pulse while black-box Krotov meets the same error at an order-of-magnitude energy premium. This method-independence is what qualifies the bounded solver as a measurement instrument, and the central result follows: a large statistical survey of unselected optimization runs resolves a weak-driving quantum speed limit for two exchange-coupled electron spins: low-amplitude realizations of the two-qubit quantum Fourier transform cease to exist below a critical gate time, and the minimum control energy diverges on approach to this limit. The divergence obeys a simple two-parameter area--pole law, $E_2^{\mathrm{law}}(T)=A/T+B/(T-T^*)$, whose first term is the time-optimal area cost, measured independently on the drift-free problem, and whose second term is a pole at the speed limit. 


\end{abstract}

\keywords{quantum optimal control, shooting methods, spin qubits, quantum speed limit, energy-efficient pulses}

\maketitle

\tableofcontents

\section{Introduction}

Quantum optimal control (QOC) underpins the realization of high-fidelity logical operations in essentially every physical platform for quantum information processing~\cite{glaser2015,koch2022,brif2010}, from superconducting circuits~\cite{krantz2019} and trapped ions~\cite{bruzewicz2019} to semiconductor spin qubits~\cite{burkard2023}  and nitrogen-vacancy centers~\cite{barry2020,tinoco:hal-05404999}. The general goal is to design a time-dependent control field that steers a finite-dimensional quantum system from the identity to a target unitary $U_{\mathrm{gate}}$ over a fixed gate duration $T$. An additional challenge arises from the presence of a constant background Hamiltonian---the drift Hamiltonian---that cannot be experimentally controlled or turned off.


Two families of algorithms have come to dominate practical QOC. Gradient ascent pulse engineering (GRAPE)~\cite{khaneja2005,defouquieres2011} discretizes the control on a time grid and updates the amplitude at each step using the analytic gradient of the gate fidelity. Krotov-type methods~\cite{krotov1996,reich2012,goerz2019} derive a monotonically convergent update rule from a Pontryagin-style variational principle and have become a standard reference for benchmarking control pulses in closed and open systems. Both families are first-order schemes for the same Pontryagin stationarity conditions; GRAPE's gradient is the Pontryagin switching function evaluated with the backward-propagated costate~\cite{khaneja2005,boscain2021}, and the fixed points of Krotov's monotonic sequential update satisfy the same first-order conditions~\cite{reich2012}; differing in the update schedule: sequential in time with a monotonicity guarantee for Krotov, concurrent over all time steps for GRAPE~\cite{machnes2011}. Both are highly flexible, but in their standard form they treat each pulse amplitude as an independent degree of freedom, and the resulting waveforms can be temporally rough and spectrally broad (parametrized formulations exist within both families~\cite{skinner2010,lucarelli2018}). Spectral breadth is benign when every available transition is dynamically useful (and each available transition far from each others), but problematic when the optimizer pours amplitude into resonances that are not central to the target gate. This increases peak amplitude, integrated pulse power, and the burden on the control electronics, at a moment when the energetic cost of control is becoming a design criterion for quantum technologies in its own right~\cite{auffeves2022,abdelhafez2020,jankovic2024,landauer1961,aifer2022}.

An alternative is to reduce the dimensionality of the control search space by writing the control field as a smooth function of a small set of meta-parameters. Examples include Chopped Random Basis  (CRAB) and dCRAB, which expand the control in a chopped random basis~\cite{caneva2011,rach2015}; GOAT, which propagates analytical gradients with respect to ansatz parameters~\cite{machnes2018}; and shooting methods inspired by the Pontryagin maximum principle (PMP)~\cite{pontryagin1962,boscain2021}. MAGICARP belongs to this last family: the control field is generated by transporting a finite-dimensional anti-Hermitian generator along the quantum trajectory and projecting it onto the available control directions~\cite{magicarp2025}, a construction recently extended to high-dimensional spin qudits~\cite{etienney2026}. The parametrization is structured rather than arbitrary: a single transported object generates all control channels, so the resulting pulses are smooth by design. The division among the three methods is thus a division of PMP costs: GRAPE and Krotov iterate toward Pontryagin stationarity of the terminal fidelity, while the shooting ansatz solves the PMP stationarity condition of the quadratic energy cost in closed form, \cref{eq:magicarp_control_law}, and reaches the target by shooting on the initial costate: it is a hybrid method where energy-optimality enters through the parametrization, fidelity through the objective.

In its original formulation, however, MAGICARP was developed and tested without drift Hamiltonian, i.e. $H_0=0$, in which the system evolution is generated solely by the control pulses. Many experimentally relevant platforms violate this assumption strongly. Static exchange-coupled spin qubits are a paradigmatic case --- in semiconductor quantum dots~\cite{loss1998,petta2005,burkard2023} as well as in the emerging atomic-scale platform of exchange-coupled spins on surfaces addressed by electron-spin-resonance scanning tunneling microscopy~\cite{Wang2023science, surfaceqb2026}: local Zeeman splittings and an always-on exchange coupling produce a drift Hamiltonian whose eigenstates are dressed (entangled) two-qubit states.
Similar drift-dominated regimes arise in dispersively coupled transmon circuits~\cite{blais2021}, NV-center registers~\cite{dutt2007}, and dipolar molecular qubits~\cite{gaitaarino2019,Biard2021}. In all of these, the free evolution generated by the drift during the gate is not a perturbation but a structural feature of the dynamics. Consequently, the control Hamiltonian must be optimized with the specific drift Hamiltonian in mind.

This work introduces a \emph{drift-aware} MAGICARP workflow for closed quantum systems with a fixed internal Hamiltonian. The control law --- a transported anti-Hermitian generator projected onto control directions --- is unchanged, but it is embedded into a two-stage pipeline. First, the full drift Hamiltonian is diagonalized to define a dressed basis (interaction-dressed throughout this work: the eigenbasis of the static drift including the exchange term, not field/control dressing of the Autler--Townes/Floquet type); the rotating-wave approximation (RWA)~\cite{shirley1965} is then performed in this basis, so that the control directions of the optimization stage are the quadratures of the actual drift-dressed transitions rather than the bare qubit operators. Second, the optimized generator is refined directly in the laboratory frame, where counter-rotating terms, off-resonant driving, and Bloch--Siegert-type shifts~\cite{blochsiegert1940} are restored automatically by exact propagation. The drift therefore enters non-perturbatively, both in the target definition $U_{\mathrm{target}} = U_{\mathrm{gate}}\, e^{-iH_0T}$ and in the propagation that transports the shooting generator.

Beyond formulating the method, the central question is physical: \emph{where does low-amplitude gate synthesis end, and at what energy cost is that boundary approached?} The comparative part (what a shooting parametrization buys over the standard gradient methods at equal fidelity) enters as the control experiment that validates the probe. Comparisons of optimizers are notoriously sensitive to their stopping rules, so we introduce a \emph{fair-halting} benchmark (Section~S2 of the Supporting Information): MAGICARP, Krotov's method, and GRAPE act on the identical model, grid, and target, and every optimization stops the moment its independently verified infidelity crosses a common threshold. At matched error, the distinguishing observable is no longer the fidelity but what each method \emph{spends} to reach them --- pulse energy, area, peak amplitude, spectral structure, and robustness. We find that the drift-aware pulses are minimal on all of these axes, that an unbiased, zero-initialized GRAPE (hereafter simply referred to as GRAPE) independently rediscovers essentially the same pulse, and that Krotov's method pays a $5$--$57\times$ energy premium. Finally, because the bounded shooting solver cannot trade energy for time, the raw statistics of a $9600$-run unselected sweep turn it into a clean probe of the \emph{weak-amplitude-driving} quantum speed limit~\cite{caneva2009,deffner2017}. For a dressed two-qubit quantum Fourier transform (QFT), low-amplitude solutions cease to exist below a sharp gate-time threshold that amplitude-unbounded GRAPE reproduces and that we identify as the time a single shared control needs to spectrally resolve the exchange-split dressed transitions. Above this threshold, the minimum control energy follows a two-parameter area-pole law whose area term is measured independently on the drift-free problem.

The paper is organized as follows. \Cref{sec:method} states the drift-control problem, reviews the MAGICARP shooting parametrization, introduces the two-stage drift-aware workflow, applies it to exchange-coupled spin qubits, and defines the control-cost metrics. \Cref{sec:exp-advantages} reports the fair-halting comparison with Krotov and GRAPE at matched verified error. \Cref{sec:magicarp-sweep} presents the unselected gate-time sweep and the speed-limit analysis. \Cref{sec:synthesis} synthesizes the findings, and \cref{sec:conclusion} presents the conclusion. The technical material is collected in the Supporting Information: the optimization algorithm (Section~S1), the fair-halting protocol (Section~S2), the threshold-robustness study at a looser halt (Section~S3), and the single-method production results (Section~S4).

\section{Drift-Aware MAGICARP}
\label{sec:method}


\subsection{Control problem and target convention}

We consider a finite-dimensional closed quantum system with Hamiltonian
\begin{equation}
H(t)
=
H_0
+
\sum_{k=1}^{N_c} u_k(t) H_k ,
\label{eq:general_hamiltonian}
\end{equation}
where $H_0$ is a time-independent drift Hamiltonian, $\{H_k\}$ are control Hamiltonians, and $u_k(t)$ are real-valued control amplitudes. The propagator satisfies $\dot U(t)=-iH(t)U(t)$ with $U(0)=\mathbb{I}$, and the goal is to implement a target quantum logical operation $U_{\mathrm{gate}}$ over a fixed duration $T$.

In a driftless control problem the target can be identified with $U_{\mathrm{gate}}$. In the presence of a drift, the drift is propagated in the equation of motion under any convention; what has to be chosen is which final unitary counts as success. Writing $U_0(T)=e^{-iH_0T}$ for the free evolution, we take
\begin{equation}
U_{\mathrm{target}}
=
U_{\mathrm{gate}}\,U_0(T).
\label{eq:drift_target_general}
\end{equation}
Equivalently, $U(T)\,U_0(T)^\dagger = U_{\mathrm{gate}}$: the drive supplies the logical operation on top of the free evolution, i.e.\ the gate is defined in the frame co-rotating with the drift. Since $H_0$ is diagonal in the dressed basis, $U_0(T)$ is a known diagonal phase matrix. Its single-qubit part is tracked as pulse-phase updates (virtual-$Z$ gates)~\cite{vandersypen2005,mckay2017}, and its exchange part is absorbed by taking the logical basis to be the dressed eigenbasis, the natural encoding for an always-on interaction~\cite{benjamin2003}. This is the convention of gate optimization with free local phases~\cite{muller2011,goerz2017}, of the DYNAMO coupled-spin benchmarks~\cite{machnes2011}, and of the optimal-control experiment of Ref.~\cite{dolde2014}, who reference the pulse phases at the start of the gate rather than at its end; the two references give the same minimum-energy law and the same gate-time threshold. The complementary laboratory-frame target~\cite{palao2003} applies where no oscillator references the phases. Here, demanding $U(T)=U_{\mathrm{gate}}$ would require the weak drive ($u/\omega\approx10^{-2}$--$10^{-1}$) to cancel a phase of order one that carries no logical content and changes with $\omega T$ at the parts-per-thousand level, adding local work that would shift the measured threshold. The form of \cref{eq:drift_target_general} follows from the frame algebra. A conjugation $U_0\,U_{\mathrm{gate}}\,U_0^\dagger$ transforms an operator between frames at one instant; a propagator connects $t=0$ to $t=T$ and carries the frame factor of each endpoint, $U_I(T)=e^{+iH_0T}\,U(T)\,e^{+iH_0\cdot0}$, where the last factor is the identity because the frames coincide at $t=0$. Requiring the co-rotating propagator to equal the gate therefore gives a one-sided composition; referencing the frames at $t=T$ instead moves the factor to the other side and gives \cref{eq:drift_target_general}. In the interaction picture \cref{eq:drift_target_general} reads $U_I(T)=U_0(T)^\dagger U_{\mathrm{gate}}\,U_0(T)$: each matrix element of the gate is multiplied by the carrier phase accumulated on its transition over the gate window, so the only freedom left is whether pulse phases are referenced to the local oscillator at the start or at the end of the gate.

The optimization loss is based on the normalized Hilbert--Schmidt (process) fidelity
\begin{equation}
\mathcal{F}(U,V)
=
\frac{1}{d^2}
\left|
\mathrm{Tr}\left(V^\dagger U\right)
\right|^2,
\qquad
\mathcal{L}
=
1-\mathcal{F}\left(U(T),U_{\mathrm{target}}\right),
\label{eq:hs_fidelity}
\end{equation}
where $d$ is the Hilbert-space dimension; the same definition is used consistently for all methods compared in this work. The average gate fidelity quoted in some references follows from $\mathcal{F}_{\mathrm{avg}} = (d\mathcal{F}+1)/(d+1)$~\cite{horodecki1999,nielsen2002}.

\subsection{Shooting parametrization}

\begin{figure}[htbp!]
  \centering
  \includegraphics[width=0.75\linewidth]{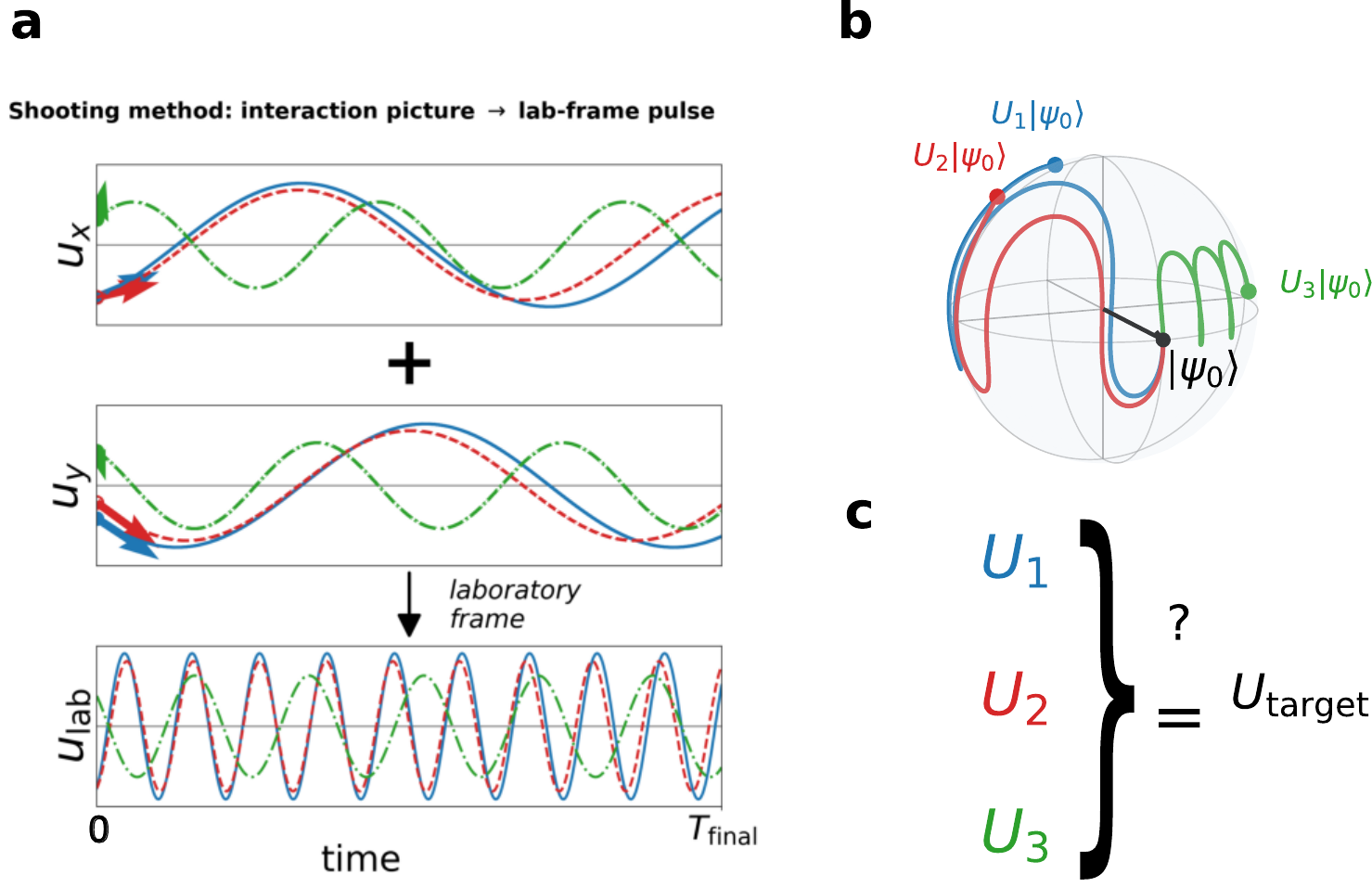}
  \caption{\textbf{The shooting (MAGICARP) construction illustrated for a single qubit.} \textbf{(a)}~A single constant generator $G(\bm\theta)$    (\cref{eq:Gtheta}) is transported along the trajectory (\cref{eq:Mtransport}), and projected onto the control quadratures to give the smooth interaction-picture controls $u_x(t)$ and $u_y(t)$ (\cref{eq:magicarp_control_law}). Re-introducing the transition frequency as a carrier recombines them into the physical laboratory pulse $u_{\mathrm{lab}}(t)$. Three different generators (blue, red, green), produce the propagators $U_1,U_2,U_3$; time runs to the gate time $T_{\mathrm{final}}$.  \textbf{(b)}~The corresponding interaction-picture evolution of an initial state $\ket{\psi_0}=(\ket{0}+i\ket{1})/\sqrt2$ on the Bloch sphere, one trajectory per generator (same colors), ending at the final states $U_k\ket{\psi_0}$. \textbf{(c)}~The optimization goal: the shooting variables $\bm\theta$ are tuned so that the generated propagator reproduces the target gate up to a global phase $\{U_1,U_2,U_3\}\stackrel{?}{\equiv}U_{\mathrm{target}}$.}
  \label{fig:shooting-schematic}
\end{figure}

MAGICARP~\cite{magicarp2025} is a shooting-based parametrization of quantum controls inspired by the PMP. Instead of optimizing each pulse amplitude independently on a time grid, the method optimizes a finite-dimensional matrix parameter --- the analogue of the initial adjoint momentum of a shooting method --- written as an anti-Hermitian generator
\begin{equation}
G(\bm{\theta})
=
\sum_{a=1}^{N_\theta}
\theta_a B_a ,
\qquad
B_a^\dagger=-B_a ,
\label{eq:Gtheta}
\end{equation}
with $\bm{\theta}\in\mathbb{R}^{N_\theta}$ the vector of optimization variables and $\{B_a\}$ an anti-Hermitian matrix basis. In the current work, the optimizer's only regularization is a box constraint $|\theta_a|\le b$ on every
component of $\bm\theta$, applied identically in both stages below; we call $b$ the
\emph{amplitude bound}. Because the generator only sets the \emph{scale} of the
reconstructed pulse [\cref{eq:magicarp_control_law,eq:lab_pulse_dressed}], $b$ carries
the units of the control amplitude ($\mathrm{rad/ns}$) but bounds $\bm\theta$, not the
laboratory pulse $u_{\mathrm{lab}}(t)$ directly. Given a propagator $U(t)$, the transported generator
\begin{equation}
M(t;\bm{\theta})
=
U(t)\,G(\bm{\theta})\,U^\dagger(t)
\label{eq:Mtransport}
\end{equation}
remains anti-Hermitian at all times, and the control amplitudes are obtained by projecting it onto the accessible Hermitian control directions,
\begin{equation}
u_k(t;\bm{\theta})
=
\Im\,\mathrm{Tr}
\left[
H_k\,M(t;\bm{\theta})
\right].
\label{eq:magicarp_control_law}
\end{equation}
Formulations with a Hermitian shooting matrix and $\Re\,\mathrm{Tr}$ are equivalent up to this notational convention \cite{magicarp2025, etienney2026}. We adopted here the anti-Hermitian convention because $G$ then belongs to the Lie
algebra $\mathfrak{u}(d)$ of the propagator group, the natural home of a
PMP costate for unitary dynamics. The transport \cref{eq:Mtransport} is therefore algebra-internal and $G$ is directly commensurate with the generators
$-iH$. Once $\bm{\theta}$ is fixed, the full waveform follows from \cref{eq:Mtransport} and~\cref{eq:magicarp_control_law}: the optimizer chooses the initial generator, and the controls follow from the trajectory that this choice generates. \Cref{eq:magicarp_control_law} should therefore not be read as a gradient-update rule: adjoint-type transport along the trajectory is indeed a hallmark of gradient-based optimal control~\cite{khaneja2005,reich2012}, but here the transported object is the decision variable itself. The method shoots on the finite-dimensional initial PMP costate $G(\bm\theta)$, and no per-sample update of $u(t)$ ever occurs. The optimizer varies $\bm{\theta}$ so that the propagator generated at the final time, $U(T)$, reproduces the target gate $U_{\mathrm{target}}$, i.e.\ so as to minimize the gate infidelity~\cref{eq:hs_fidelity}. \Cref{fig:shooting-schematic} illustrates this construction on a single qubit.

The construction restricts the search space compared with fully time--discretized methods such as GRAPE or Krotov, but the restriction is structured: all control channels are generated from the same transported object, which correlates their time dependence and yields smooth pulses whenever the propagator varies smoothly. The parameter count is likewise fixed by the generator alone, independent of the number of control channels and of the time grid: a multi-channel dressed RWA stage ($2N_{\mathrm{transitions}}$ quadrature controls) therefore costs the shooting method nothing, while for pixelwise methods every added channel multiplies the number of decision variables. This control-count independence is what makes the parametrization naturally suited to the two-stage drift-aware workflow. In the present work the algebraic control law is unchanged with respect to the driftless formulation; what changes is that the propagator transporting $G(\bm{\theta})$ is now shaped by the drift. 

\subsection{Two-stage workflow: dressed-frame initialization and laboratory-frame refinement}
\label{sec:workflow}

\begin{figure}[htbp!]
\centering
\includegraphics[width=\linewidth]{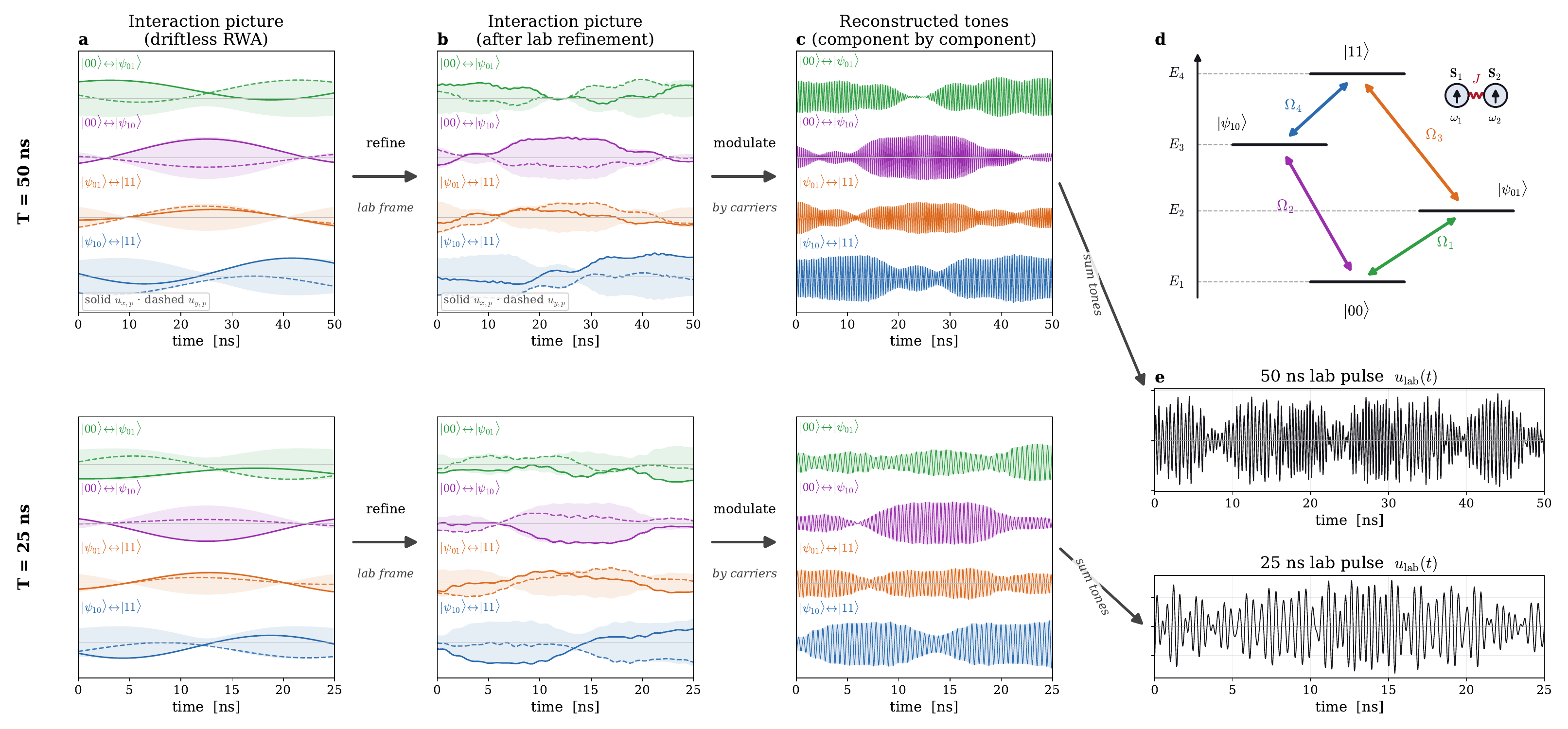}
\caption{\textbf{Drift-aware dressed-transition shooting workflow}, shown for the dressed
two-qubit QFT, scalar control
$H_{\mathrm{c}}=X_1+X_2$), at gate times $T=50\,\mathrm{ns}$ (top rows) and $T=25\,\mathrm{ns}$
(bottom rows). A single anti-Hermitian generator $G(\bm\theta)$ [\cref{eq:Gtheta}], obtained after regular MAGICARP optimization in the RWA, is
transported along the trajectory and projected onto the dressed-transition quadratures
[\cref{eq:magicarp_control_law,eq:dressed_quadratures}].
\textbf{(a)}~Interaction-picture quadratures $u_{x,p}(t)$ (solid) and $u_{y,p}(t)$ (dashed),
one complex envelope per dressed transition, from the driftless rotating-wave stage
[\cref{eq:HRWA_dressed}]; no counter-rotating terms.
\textbf{(b)}~The same quadratures after the exact laboratory-frame refinement of $\bm\theta$;
the slow envelopes are preserved but now carry the fast drift (counter-rotating) corrections.
\textbf{(c)}~Each refined pair modulated onto its dressed-transition carrier,
$2[u_{x,p}\cos\Omega_p t-u_{y,p}\sin\Omega_p t]$, reconstructed transition by transition
[\cref{eq:lab_pulse_dressed}].
\textbf{(d)}~Dressed levels $E_1$--$E_4$ of the full drift $H_0$ [\cref{eq:spin_full_drift}]
and the four control-coupled transitions $\Omega_1$--$\Omega_4$ (inset: the two exchange-coupled
spins, exchange $J$).
\textbf{(e)}~Their sum is the scalar laboratory pulse
$u_{\mathrm{lab}}(t)=\sum_p 2[u_{x,p}\cos\Omega_p t-u_{y,p}\sin\Omega_p t]$ [\cref{eq:lab_pulse_dressed}]
applied to the device; both gate times reach $1-\mathcal{F}\sim10^{-14}$ (Table~S6 of the Supporting Information).
Colour code: $\Omega_1=\ket{00}\!\leftrightarrow\!\ket{\psi_{01}}$ (green),
$\Omega_2=\ket{00}\!\leftrightarrow\!\ket{\psi_{10}}$ (purple),
$\Omega_3=\ket{\psi_{01}}\!\leftrightarrow\!\ket{11}$ (orange),
$\Omega_4=\ket{\psi_{10}}\!\leftrightarrow\!\ket{11}$ (blue).}
\label{fig:workflow}
\end{figure}

The workflow rests on a simple strategy: first solve an \emph{approximate}
but structurally transparent version of the control problem --- the dynamics
in the dressed rotating frame under the RWA --- to obtain a good candidate
generator $\bm\theta$, and then inject this candidate as the initialization
of the \emph{exact} laboratory-frame problem. The solution of this stage is the one relevant to the control optimization in the presence of the drift Hamiltonian, see~\cref{fig:workflow}.

The laboratory-frame Hamiltonian is written with a single scalar control channel,
\begin{equation}
H_{\mathrm{lab}}(t)
=
H_0
+
u_{\mathrm{lab}}(t)\,H_{\mathrm{c}},
\qquad
H_0 = H_{\mathrm{local}} + H_{\mathrm{cpl}},
\label{eq:lab_hamiltonian_scalar}
\end{equation}
where $H_0$ is the \emph{full} drift (local terms $H_{\mathrm{local}}$ plus coupled terms $H_{\mathrm{cpl}}$) and $H_{\mathrm{c}}$ the laboratory control operator. The first optimization stage is constructed in the interaction frame of the full drift --- not in a local rotating frame in which the coupling would survive as a residual time-dependent term. Diagonalizing $H_0\ket{\varepsilon_i}=E_i\ket{\varepsilon_i}$ defines the dressed basis. Because the coupling is always on, these eigenstates are not the bare product
states but superpositions \emph{dressed} by $H_{\mathrm{cpl}}$: for the
exchange-coupled pair below $\ket{00}$ and $\ket{11}$ remain (nearly) product
states while $\ket{01}$ and $\ket{10}$ hybridize into the entangled
single-excitation pair $\ket{\psi_{01}}$ and $\ket{\psi_{10}}$ \cite{surfaceqb2026}.
These are the states the system follows between pulses, so both the logical
basis and the drive frequencies must be referred to them. The dressed levels and the four control-coupled transitions are shown in \cref{fig:workflow}d. Each pair of dressed states $p=(i,j)$, $i<j$, coupled by the control ($s_p=\bra{\varepsilon_j}H_{\mathrm{c}}\ket{\varepsilon_i}\neq0$) defines a retained transition with frequency $\Omega_p=E_j-E_i$, transition operator $A_p=s_p\ket{\varepsilon_j}\bra{\varepsilon_i}$, and Hermitian quadratures
\begin{equation}
H_{x,p}
=
A_p+A_p^\dagger,
\qquad
H_{y,p}
=
i\left(A_p^\dagger-A_p\right).
\label{eq:dressed_quadratures}
\end{equation}
The RWA control directions are these dressed-transition quadratures $\{H_{x,p},H_{y,p}\}_{p\in\mathcal{T}}$, so the exchange coupling enters the RWA stage non-perturbatively through the dressed eigenstates, transition frequencies, and matrix elements.

In the dressed RWA stage the transported generator is $M=U_I\,G(\bm\theta)\,U_I^\dagger$ with $U_I$ the interaction-frame propagator, the channel amplitudes are $u_{x,p}=\Im\,\mathrm{Tr}[H_{x,p}M]$ and $u_{y,p}=\Im\,\mathrm{Tr}[H_{y,p}M]$, and the propagation Hamiltonian is
\begin{equation}
H_{\mathrm{RWA}}(t;\bm{\theta})
=
\sum_{p\in\mathcal{T}}
\left[
u_{x,p}(t;\bm{\theta})H_{x,p}
+
u_{y,p}(t;\bm{\theta})H_{y,p}
\right],
\label{eq:HRWA_dressed}
\end{equation}
integrated stepwise by matrix exponentials. The RWA-stage loss compares $e^{-iH_0T}U_I(T;\bm\theta)$ with $U_{\mathrm{target}}$, and its optimizer $\bm\theta_{\mathrm{RWA}}$ initializes the second stage.

The second stage refines $\bm\theta$ under the exact laboratory dynamics. At each time step the interaction-frame propagator $U_I(t)=e^{+iH_0t}U_{\mathrm{lab}}(t)$ transports the generator, the same dressed quadrature amplitudes $u_{x,p},u_{y,p}$ are evaluated, and the physical scalar pulse is reconstructed as a sum over carriers at the dressed transition frequencies:
\begin{equation}
u_{\mathrm{lab}}(t;\bm{\theta})
=
\sum_{p\in\mathcal{T}}
\alpha_p
\left[
u_{x,p}(t;\bm{\theta})\cos(\Omega_p t+\phi_p)
-
u_{y,p}(t;\bm{\theta})\sin(\Omega_p t+\phi_p)
\right],
\label{eq:lab_pulse_dressed}
\end{equation}
with $\alpha_p=2$ and $\phi_p=0$ unless stated otherwise. \Cref{eq:lab_pulse_dressed} makes the relation between the two stages explicit: the RWA stage treats the slowly varying quadratures as effective controls acting directly on the dressed transitions (keeping the resonant terms generated by the carriers at $\Omega_p$), while the laboratory-frame refinement propagates the full Hamiltonian $H_0+u_{\mathrm{lab}}(t)H_{\mathrm{c}}$ with the explicitly oscillating pulse, so that counter-rotating terms, off-resonant driving, and Bloch--Siegert corrections~\cite{blochsiegert1940} are included automatically. The final fidelity and all pulse diagnostics are evaluated in this exact laboratory frame; the RWA stage serves only as a structured initialization. Two distinct approximations enter that initialization. The single-tone rotating-wave approximation, which neglects the counter-rotating terms of each carrier, requires the quadrature amplitudes to be small against the transition frequencies $\Omega_p$; the secular approximation, which treats the retained transitions $p\in\mathcal{T}$ as independently addressable channels (\cref{eq:lab_pulse_dressed}), requires in addition that the pulse's spectral width $\sim2\pi/T$ be small against the spacing of the transition frequencies~\cite{vandersypen2005}. For the spin model of \cref{sec:spinqb} the two conditional transitions of each qubit are split by the exchange $J$ and the two qubit groups by $\omega_1-\omega_2$~\cite{zajac2018,russ2018}, so this assumption holds only for $T\gtrsim2\pi/J+2\pi/(\omega_1-\omega_2)\approx12\,\mathrm{ns}$. Below that scale the dressed stage ceases to be a faithful initialization; because the refinement propagates the exact Hamiltonian, the failure appears as verified laboratory-frame infidelity rather than as an unnoticed error, and \cref{sec:magicarp-sweep} measures where the weak-driving optimum actually leaves this domain. The complete two-stage algorithm, including the L-BFGS-B optimization~\cite{byrd1995}, amplitude bounds, and multistart strategy, is given in Section~S1 of the Supporting Information.

\subsection{Exchange-coupled spin qubits and operating points}
\label{sec:spinqb}

As a representative application we consider two exchange-coupled spin-$\tfrac12$ qubits~\cite{loss1998,petta2005,burkard2023,surfaceqb2026}. The  drift Hamiltonian is given by
\begin{equation}
H_0=-\omega_1 Z_1 - \omega_2 Z_2 + J\,\mathbf{S}_1\!\cdot\!\mathbf{S}_2 ,
\qquad
\mathbf{S}=\tfrac{\boldsymbol\sigma}{2},
\label{eq:spin_full_drift}
\end{equation}
where $\omega_i$ are the on-site energies (Larmor frequencies), $J$ is the Heisenberg exchange coupling strength and $\mathbf{S}=(X, Y, Z)$ is the vector of spin operators.  The scalar laboratory control is given by
\begin{equation}
H_{\mathrm{c}}
=
X_1+X_2 .
\label{eq:spin_lab_control}
\end{equation}
The full drift --- not only the local Zeeman part --- is diagonalized to define the dressed basis, consistent with the exchange being static and always on during the gate. Logical gates are also specified in the dressed basis as it represents the states the system actually follows in the absence of control. The targets considered are the $\mathrm{NOT}_2$ gate, $\mathrm{NOT}_2\equiv I\otimes X$, the single-qubit bit flip of the second logical qubit (dressed-basis form $B\,(I\otimes X)\,B^\dagger$; the name $\mathrm{NOT}_2$ follows Ref.~\cite{surfaceqb2026}), and the fully entangling two-qubit quantum Fourier transform $U_{\mathrm{QFT}}^{\mathrm{dressed}}=B\,F_4\,B^\dagger$, with $B$ the dressed-eigenbasis matrix and $F_4$ the four-dimensional Fourier matrix~\cite{nielsenchuang}.

We consider two operating points (see Tab.~\ref{tab:operating-points} for a list of parameters used) which differ in their relative coupling strengths $J$ and hence represent two different drift Hamiltonians:

\begin{enumerate}
    \item A \emph{strong-coupling} regime which reproduces the benchmark study of Ref.~\cite{surfaceqb2026}. Here, $\omega_1=20 \pi\,\mathrm{GHz}$, $\omega_2=14\pi\,\mathrm{GHz}$, isotropic exchange $J=5\,\mathrm{GHz}$, with the four dressed transitions at $\Omega_p/2\pi\in\{6.55,\,7.35,\,9.65,\,10.45\}\,\mathrm{GHz}$; the $\mathrm{NOT}_2$ action is carried by the $\ket{00}\!\leftrightarrow\!\ket{\psi_{01}}$ and $\ket{\psi_{10}}\!\leftrightarrow\!\ket{11}$ transitions ($6.55$ and $7.35\,\mathrm{GHz}$). The other two transitions being spectators.
    \item A \emph{moderate-coupling} point which uses $\omega_1=17\,\mathrm{GHz}$, $\omega_2=13\,\mathrm{GHz}$ and $J=0.6\,\mathrm{GHz}$; The smaller $J/(\omega_1-\omega_2)$ here leads to a weaker hybridization of the eigenstates. The two gate-essential $\mathrm{NOT}_2$ transitions sit near the bare qubit-2 frequency ($13/2\pi =2.06$ GHz, $\Omega_p/2\pi\approx2.02$ and $2.11\,\mathrm{GHz}$) and the two spectators near qubit-1 ($17/2\pi =2.70$ GHz, $\Omega_p/2\pi\approx 2.66$ and $2.76\,\mathrm{GHz}$).
\end{enumerate}

For both cases we chose gate times between $T=12.5$ ns and $T=50$ ns. The time steps used are  $\Delta t=0.01\,\mathrm{ns}$ and $\Delta t=0.02\,\mathrm{ns}$, respectively. Throughout this work, times are in ns, frequencies in GHz, and the control amplitude $u_{\mathrm{lab}}$ in $\mathrm{rad/ns}$, so that the pulse energy $E_2$ carries units of $\mathrm{rad^2/ns}$ and the pulse area $E_1$ of rad (\cref{sec:metrics}).


\begin{table}[htbp]
\centering
\begin{threeparttable}
\caption{Summary of the two operating points used as testbeds. Case~(i) is the
strongly coupled regime; case~(ii) is the moderate-exchange point.
Larmor frequencies $\omega_i$ and exchange-coupling constants $J$ are in GHz; the four
dressed transitions $\Omega/2\pi$ separate into two gate-essential $\mathrm{NOT}_2$
transitions and two spectators.}
\label{tab:operating-points}
\begin{tabular}{@{}lll@{}}
\toprule
Quantity & (i) Strong coupling  &  (ii) Moderate coupling  \\
\midrule
$\omega_1$ (GHz) & $20\pi$ & $ 17$  \\
$\omega_2$ (GHz) & $14\pi$ & $13$  \\
$J$ (GHz) & $5$& $0.6$  \\
$J/(\omega_1-\omega_2)$ & $\approx 0.3$ & $\approx 0.15$ \\
Dressed transitions $\Omega_p/2\pi$ (GHz)& $6.55,\ 7.35,\ 9.65,\ 10.45$ & $2.02,\ 2.11,\ 2.66,\ 2.76$ \\
Gate-essential ($\mathrm{NOT}_2$) (GHz) & $6.55,\ 7.35$ & $2.02,\ 2.11$  \\
Spectators (GHz) & $9.65,\ 10.45$ & $2.66,\ 2.76$  \\
Gate times $T$ (ns)& $\{50,\ 30,\ 12.5\}$\tnote{a} & $\{50,\ 25,\ 12.5\}$  \\
\bottomrule
\end{tabular}

\begin{tablenotes}\small
\item[a] See Section~S3.2 of the Supporting Information
\end{tablenotes}

\end{threeparttable}
\end{table}

\subsection{Control-cost metrics}
\label{sec:metrics}

The optimization objective is the gate infidelity alone; energy efficiency and spectral parsimony (a pulse being spectrally parsimonious when its power concentrates on the gate-essential dressed transitions and (almost) nowhere else, quantified below by the gate-tone spectral fraction) are evaluated \textit{a posteriori} from the optimized laboratory pulse. The integrated pulse power
\begin{equation}
E_2
=
\int_0^T
|u_{\mathrm{lab}}(t)|^2\,dt
\label{eq:E2}
\end{equation}
is the most direct proxy for the power delivered to the device~\cite{abdelhafez2020,auffeves2022}; by Parseval's theorem it equals, up to normalization, the integrated spectral power. $u_{\mathrm{lab}}$ is an angular Rabi rate ($\mathrm{rad/ns}$), so $E_2$ carries $\mathrm{rad^2/ns}$ hence $\hbar E_2$ has units of energy; $E_2$ is proportional to the electromagnetic energy delivered to the control line. This places our energy metrics in the broader context of the thermodynamic cost of computation~\cite{landauer1961} and recent connections between speed limits and energy-efficient quantum gates~\cite{aifer2022}. This is relevant as the optimization can distribute amplitude directly in the driving pulses or elsewhere on the frequency axis. In the latter case, this  parasitic spectral weight translates directly into control energy. We also report the pulse area
\begin{equation}
E_1=\int_0^T|u_{\mathrm{lab}}(t)|\,dt ,
\label{eq:E1}
\end{equation}
fixed, in the weak-driving regime, by the rotation angles the gate must execute and therefore expected to be conserved across gate times, and the peak amplitude $\|u_{\mathrm{lab}}\|_\infty$, which the control electronics must support. Spectral structure is quantified by the \emph{gate-tone spectral fraction}: the share of squared spectral weight $|\tilde u(f)|^2$ located at the gate-essential dressed transitions versus the spectator transitions, evaluated by peak-integrated, data-driven windows (Section~S2 of the Supporting Information). For a $\mathrm{NOT}_2$ gate only two dressed transitions are dynamically central, so a parsimonious pulse should be two-tone; for the QFT the useful set is the full dressed manifold.


\section{Advantages of MAGICARP from an Experimental Point of View}
\label{sec:exp-advantages}

All three optimizers compared in this work can be driven to essentially the same
fidelity; what an experiment must actually implement is not the abstract generator but the physical pulse. An arbitrary-waveform generator has finite bandwidth and amplitude
range, spectator transitions are driven by whatever power the pulse leaks off
the gate-essential resonances, and the device parameters the pulse was optimized
for are never known exactly. In this section we therefore compare, at equal
verified gate infidelity, the four observables that decide whether a pulse is
implementable. These are the pulse's

\begin{enumerate}
    \item spectral content (\cref{sec:spectral}),
    \item energy,
area, and peak-amplitude cost (\cref{sec:pathological}), and
\item robustness
to parameter error (\cref{sec:robustness}), and
\item the temporal resolution it demands of the waveform generator (\cref{sec:spectral}, Figure~S1 of the Supporting Information).
\end{enumerate}

These axes separate the optimization methods sharply where fidelity alone does not, and they set up the speed-limit analysis of \cref{sec:magicarp-sweep}.

Throughout this section we work at the moderate-exchange operating point of
\cref{sec:spinqb},$\omega_1 = 17\,\mathrm{GHz}$, $\omega_2 = 13\,\mathrm{GHz}$ with the $\mathrm{NOT}_2$ and dressed-QFT targets,
gate times $T \in \{50, 25, 12.5\}\,\mathrm{ns}$, and
$\Delta t = 0.02\,\mathrm{ns}$. The strongly coupled
surface-qubit benchmark regime~\cite{surfaceqb2026} is treated in the
Supporting Information (Sections~S2--S4).

The comparisons follow a \emph{fair-halting} protocol, given in full in
Section~S2 of the Supporting Information: MAGICARP, Krotov's method, and zero-initialized
GRAPE act on identical models, grid, and dressed target, and every
optimization stops the moment its \emph{verified} infidelity (independent
re-propagation of the stored pulse) crosses $1-\mathcal{F}\le10^{-5}$, or when
its $1000$-iteration budget is exhausted. Each method runs a four-member
multistart; among converged restarts the lowest-energy pulse is kept, and if
none converge the lowest-infidelity run is reported and flagged with an
asterisk. Beyond a satisfactory fidelity, the comparison is about cost, not capability.

\subsection{Spectrally concentrated pulses}
\label{sec:spectral}

\begin{figure}[htbp]
\centering\includegraphics[width=\linewidth]{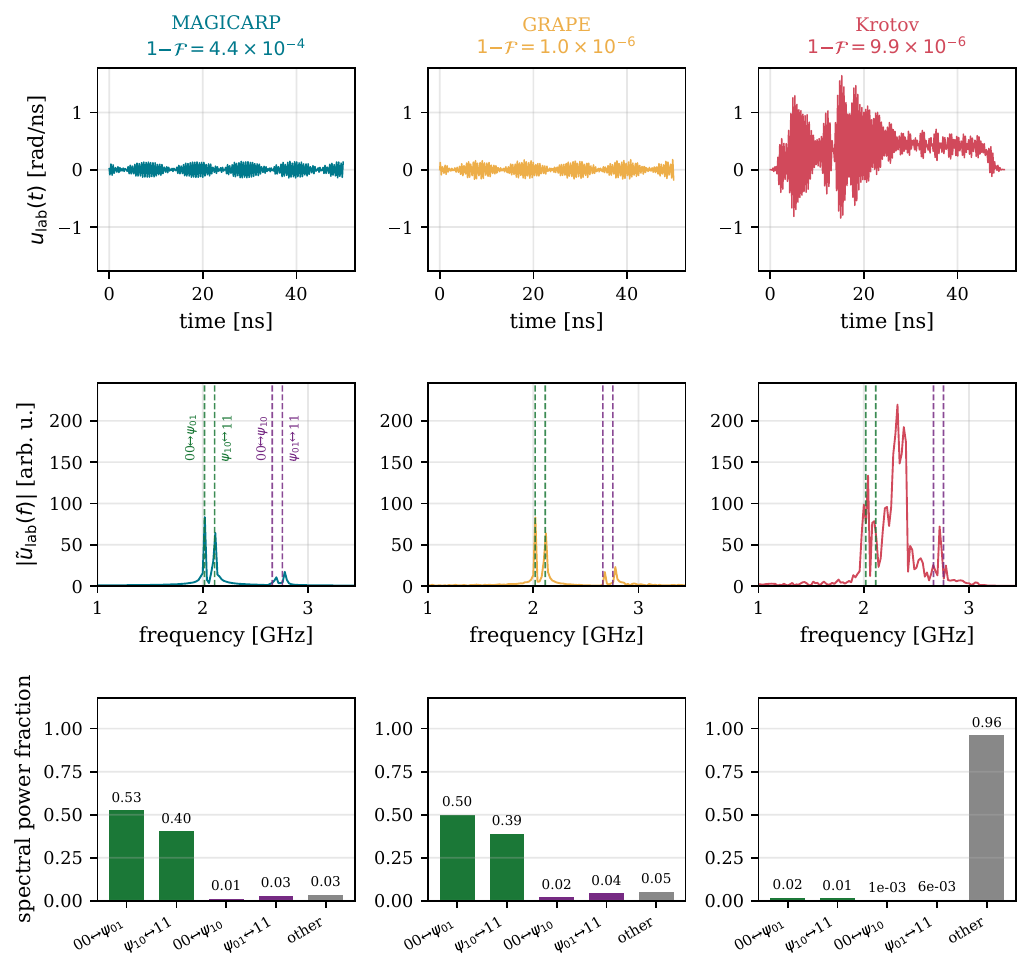}
\caption{Anatomy of the $\mathrm{NOT}_2$ pulses at moderate exchange, $T=50\,\mathrm{ns}$,
fair $10^{-5}$ halt. Columns: MAGICARP, GRAPE, Krotov. Top: laboratory pulse
$u_{\mathrm{lab}}(t)$ (shared vertical scale). Middle: spectrum
$|\tilde u_{\mathrm{lab}}(f)|$, with the two gate-essential (green) and two
spectator (purple) dressed transitions marked; the vertical scale is set by the
transition band, so the Krotov pulse's large zero-frequency component runs off
scale. Bottom: share of total spectral power integrated over each transition
peak (peak windows from a \texttt{find\_peaks}/\texttt{peak\_widths} pass on
$|\tilde u|^2$~\cite{virtanen2020}; asterisks mark transitions where no peak was
detected and a fixed $\pm2$-bin window was used); the grey bar collects all
off-transition power. The MAGICARP and GRAPE pulses are smooth, low-amplitude,
and two-tone ($93\%$ and $88\%$ of their power on the gate transitions); the
Krotov pulse is broadband, with $96\%$ of its power off every dressed
transition.}
\label{fig:not2-spec3x3}
\end{figure}

\begin{figure}[htbp]
\centering\includegraphics[width=\linewidth]{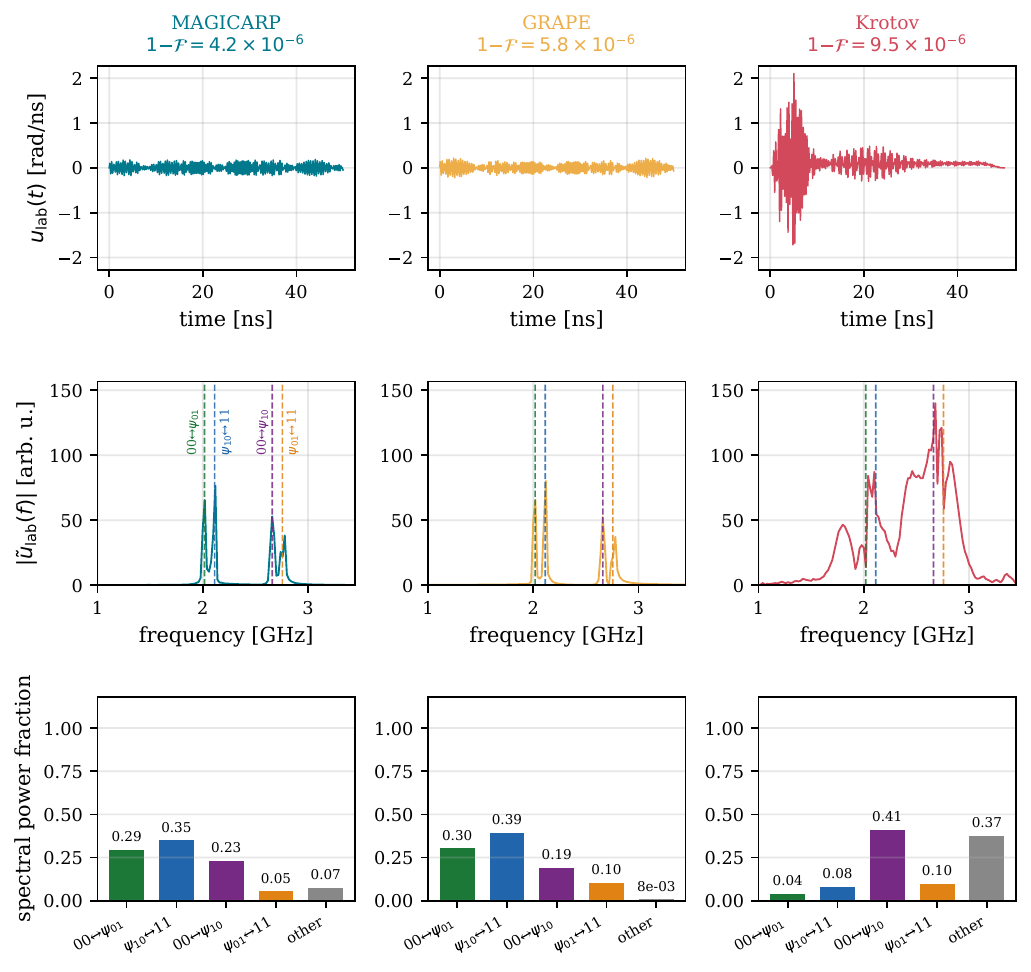}
\caption{Same anatomy for the dressed QFT at $T=50\,\mathrm{ns}$, fair $10^{-5}$ halt.
Because the QFT is fully entangling, its useful set is the entire dressed
manifold: the bottom row shows the spectral power distributed over all four
dressed transitions (one color per transition). MAGICARP and GRAPE realize the
four-tone structure at minimal energy ($93\%$ and $99\%$ of their power on the
four transitions; $E_2=0.354$ vs $0.361\,\mathrm{rad^2/ns}$); Krotov satisfies the same error with
$14\times$ the energy, $38\%$ of which lies off every transition, concentrated
in an uneven broadband hump.}
\label{fig:qft-spec3x3}
\end{figure}

Spectrally concentrated pulses are preferable for experimental implementation: they
fit within finite electronics bandwidth, they do not drive spectator
transitions, and they are physically interpretable as Rabi drive on identified
dressed lines. We use \emph{smooth} throughout in an operational rather than a differentiability sense: a smooth pulse is band-limited, free of step discontinuities, and free of any slew rate beyond what the generator can realize between consecutive samples. What this buys is that the optimized pulse and the emitted pulse are the same object. A waveform whose content reaches the generator's Nyquist frequency cannot be written to the grid faithfully: truncating it there produces Gibbs ringing at every fast edge, with the characteristic overshoot inflating the peak amplitude the electronics must supply, and the reconstruction images fold back into the band as broadband spurs. Those spurs are the expensive part, because a pulse designed to address two dressed transitions then deposits power within a linewidth of the other two, reintroducing exactly the spectator driving and leakage that the spectral concentration was meant to remove. An envelope confined to a few times $1/T$ has no content anywhere near the sampling resolution, so neither artifact arises, and smoothness in this sense is a consequence of the construction rather than a prior imposed on the solution. By construction --- \cref{eq:lab_pulse_dressed} --- MAGICARP
emits carriers only at the dressed transition frequencies; the question this
subsection answers is what the unconstrained methods emit at the same verified
error. \Cref{fig:not2-spec3x3,fig:qft-spec3x3} show, for each
method, the converged $T=50\,\mathrm{ns}$ laboratory pulse, its spectrum, and the share of
total spectral power $|\tilde u(f)|^2$ attributed to each dressed transition,
with the per-transition integration windows determined by a peak-detection pass
on the power spectrum 
(Section~S2 of the Supporting Information, reported observables). The same representation settles the fourth axis of the list above, the temporal resolution the pulse demands of the waveform generator. Because \cref{eq:lab_pulse_dressed} presents the pulse as slow envelopes on fixed carriers, an IQ-modulated source can emit it at $0.4\,\mathrm{GS/s}$, against the $5.5\,\mathrm{GS/s}$ that direct synthesis of the $2$--$3\,\mathrm{GHz}$ waveform demands of all three methods alike; the broadband pulse needs more even on the envelope route. That study reuses these same converged pulses and is reported with them in the Supporting Information (Section~S3.1, generator temporal resolution, and Figure~S1).

\paragraph{$\mathrm{NOT}_2$: two-tone concentration.} The optimized MAGICARP
pulse places most of its weight on the two gate-essential dressed transitions
and only weakly excites the spectators (\cref{fig:not2-spec3x3}): at $T=50\,\mathrm{ns}$
the peak-integrated power fractions are $0.93$ on the gate pair against $0.04$
on the spectator pair. The spectator weight is not noise --- it is the small
coherent drive needed to cancel the off-resonant cross-talk, and it grows as
the gate and spectator transitions get spectrally closer.

\paragraph{Krotov is broadband.} At matched error the Krotov pulse is spectrally unstructured: the bottom row of \cref{fig:not2-spec3x3} assigns $96\%$ of its power to no dressed transition at all, against only $4\%$ off-transition for MAGICARP and GRAPE. The flat-top-seeded Krotov flow --- as used, for example, in Ref.~\cite{surfaceqb2026} --- settles in a high-energy, broadband region of the landscape that satisfies the fidelity constraint without ever acquiring the resonant two-tone structure; the QFT energy premium this entails is quantified in \cref{sec:pathological}. In the strong-coupling regime the contrast sharpens: at the $10^{-5}$ halt the MAGICARP pulse suppresses spectator weight to $8\times10^{-5}$ of the total, two orders of magnitude below even the converged GRAPE pulse at equal energy (Section~S2 of the Supporting Information); at the looser halt of Section~S3 of the Supporting Information, however, GRAPE is equally clean, so the structural suppression is intrinsic to MAGICARP while the residual leakage is halt-dependent. \Cref{sec:magicarp-sweep} shows that this minimum-energy branch is also independent of MAGICARP's own amplitude bound.

\paragraph{The QFT uses all dressed transitions.} Spectral parsimony is not an
imposed smoothness prior --- it is gate-dependent, and the dressed QFT is its
natural counterpoint. Unlike the two-tone $\mathrm{NOT}_2$ pulse, the QFT
drives all four dressed transitions with comparable amplitude and produces a
four-peak spectrum (\cref{fig:qft-spec3x3}): the ``useful set'' of
transitions for a fully entangling gate is the entire dressed-transition
manifold, and the optimizer correctly populates every resonance. The
energy--time trade-off is preserved ($E_2$ and the peak amplitude grow as the
gate is compressed, \cref{fig:fair-metrics-vsT}), while the pulse area
$E_1\approx3.5\,\mathrm{rad}$ is roughly conserved and is larger than for $\mathrm{NOT}_2$
($\approx2.7\,\mathrm{rad}$), consistent with the QFT being a ``larger'' rotation. The
bottom row of \cref{fig:qft-spec3x3} again separates the methods: MAGICARP
and GRAPE place $93\%$ and $99\%$ of their power on the four transitions,
whereas the Krotov pulse leaves $38\%$ of a total power an order of magnitude larger off every dressed line.

\subsection{Pathological pulses cheat past the first quantum speed limit}
\label{sec:pathological}

Read naively, a convergence table says that Krotov outperforms other methods at a gate-time
limit of $T=12.5\,\mathrm{ns}$: it is the only method that
reaches the $10^{-5}$ halt there. However it is not sufficient to look at infidelity alone. The converged short-$T$ Krotov pulse is pathological: its energy, area, and peak amplitude sit an order of magnitude above the minimal pulse family, so it leaves the \emph{weak-driving regime} --- the regime $\|u_{\mathrm{lab}}\|_\infty<\Omega_p$ in which the drive amplitude is small compared with the dressed transition frequencies and the dressed-carrier control model (and any realistic drive line) is accurate. Whereas the minimal-energy pulses keep $\|u_{\mathrm{lab}}\|_\infty\sim0.2$--$0.4\,\mathrm{rad/ns}$, i.e.\ $\|u_{\mathrm{lab}}\|_\infty/\Omega_p\sim10^{-2}$, the converged $T=12.5\,\mathrm{ns}$ Krotov pulse reaches $\|u_{\mathrm{lab}}\|_\infty\approx6.2\,\mathrm{rad/ns}$, comparable to $\Omega_p\approx12.7$--$17.3\,\mathrm{rad/ns}$ ($\|u_{\mathrm{lab}}\|_\infty/\Omega_p\approx0.4$--$0.5$): the weak-driving inequality is plainly violated. The first, weak-amplitude quantum speed limit is real, and an unconstrained optimizer passes it only by leaving the physically preferred pulse class that could be detrimental for the stability of the system for example.

\begin{figure}[htbp]
\centering\includegraphics[width=\linewidth]{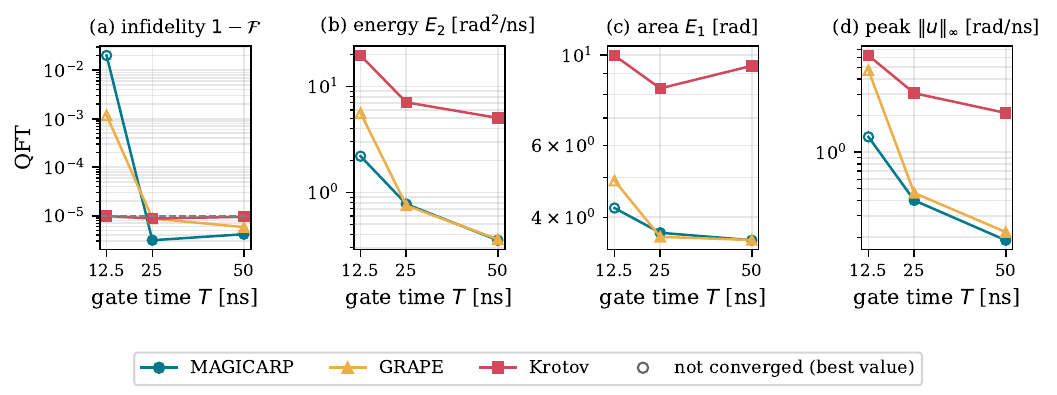}
\caption{Fair-halting comparison at the verified $10^{-5}$ halt for the dressed QFT: (a) verified infidelity (dashed line: halt target), (b) pulse energy $E_2$, (c) pulse area $E_1$, and (d) peak amplitude versus gate time. Open markers denote runs that exhausted their budget without converging (best value shown). GRAPE tracks the MAGICARP energy, area, and amplitude curves almost exactly; Krotov meets the same error criterion with an order of magnitude more energy, a non-conserved area, and up to $\sim\!10\times$ the peak amplitude.}
\label{fig:fair-metrics-vsT}
\end{figure}

\paragraph{Four metrics versus gate time.} Under the fair-halting protocol the
dressed QFT separates the three methods by \emph{control cost}, not by
infidelity (\cref{tab:qft-fair}, \cref{fig:fair-metrics-vsT}). At
$T=50\,\mathrm{ns}$ and $T=25\,\mathrm{ns}$ all three optimizers reach the target error, but MAGICARP and
GRAPE converge onto essentially the \emph{same} minimal-energy solution: their
energies agree to within $2$--$3\%$ ($E_2 = 0.354$ vs $0.361\,\mathrm{rad^2/ns}$ at $T=50\,\mathrm{ns}$;
$0.776$ vs $0.753$ at $T=25\,\mathrm{ns}$), as do their areas and peak amplitudes. The
GRAPE pulse using zero initialization which is free of any frequency preference, independently rediscovers the four-tone, low-energy pulse that the dressed parametrization produces by
construction. Krotov, climbing from a flat-top guess, satisfies the same error
criterion with $14\times$ ($T=50\,\mathrm{ns}$) and $9\times$ ($T=25\,\mathrm{ns}$) more pulse energy, $\sim\!10\times$ the peak amplitude, and $2$--$3\times$ the pulse area. The
area panel of \cref{fig:fair-metrics-vsT} makes the structural difference
visible at a glance: for MAGICARP and GRAPE the area is conserved across gate
times ($E_1\approx3.5\,\mathrm{rad}$ for the QFT, fixed by the rotation the gate implements) while energy and peak amplitude rise as
$\sim 1/T$; Krotov's area is $3$--$7\times$ larger and not conserved, the
signature of a pulse that is not a minimal realization of the rotation.

\begin{table}[htbp]
\centering
\begin{threeparttable}
\caption{Dressed QFT under the fair $10^{-5}$ halt: verified infidelity, pulse
energy $E_2$, pulse area $E_1$, and peak amplitude for MAGICARP, Krotov, and
GRAPE.}
\label{tab:qft-fair}
\begin{tabular}{llcccc}
\toprule
$T$ (ns) & method & $1-\mathcal{F}$ & $E_2$ & $E_1$ & $\|u_{\mathrm{lab}}\|_\infty$ \\
\midrule
50   & MAGICARP & $4.2\times10^{-6}$ & 0.354 & 3.51 & 0.19 \\
     & Krotov   & $9.5\times10^{-6}$ & 5.04  & 9.41 & 2.10 \\
     & GRAPE    & $5.8\times10^{-6}$ & 0.361 & 3.52 & 0.22 \\
\midrule
25   & MAGICARP & $3.1\times10^{-6}$ & 0.776 & 3.66 & 0.40 \\
     & Krotov   & $8.7\times10^{-6}$ & 7.08  & 8.28 & 3.06 \\
     & GRAPE    & $8.8\times10^{-6}$ & 0.753 & 3.58 & 0.46 \\
\midrule
12.5 & MAGICARP & $2.0\times10^{-2}$\tnote{*} & 2.20 & 4.21 & 1.34 \\
     & Krotov   & $9.9\times10^{-6}$ & 19.6  & 9.97 & 6.25 \\
     & GRAPE    & $1.2\times10^{-3}$\tnote{*} & 5.61 & 4.92 & 4.73 \\
\bottomrule
\end{tabular}

\begin{tablenotes}\small
\item[*] Runs that exhausted their budget without reaching $10^{-5}$ (best value reported).
\item[] Note: Here and in all following tables, $E_2$ (\cref{eq:E2}) is in $\mathrm{rad^2/ns}$, $E_1$ (\cref{eq:E1}) in rad, and $\|u_{\mathrm{lab}}\|_\infty$ in rad/ns.
\end{tablenotes}

\end{threeparttable}
\end{table}


\begin{figure}[htbp]
\centering\includegraphics[width=\linewidth]{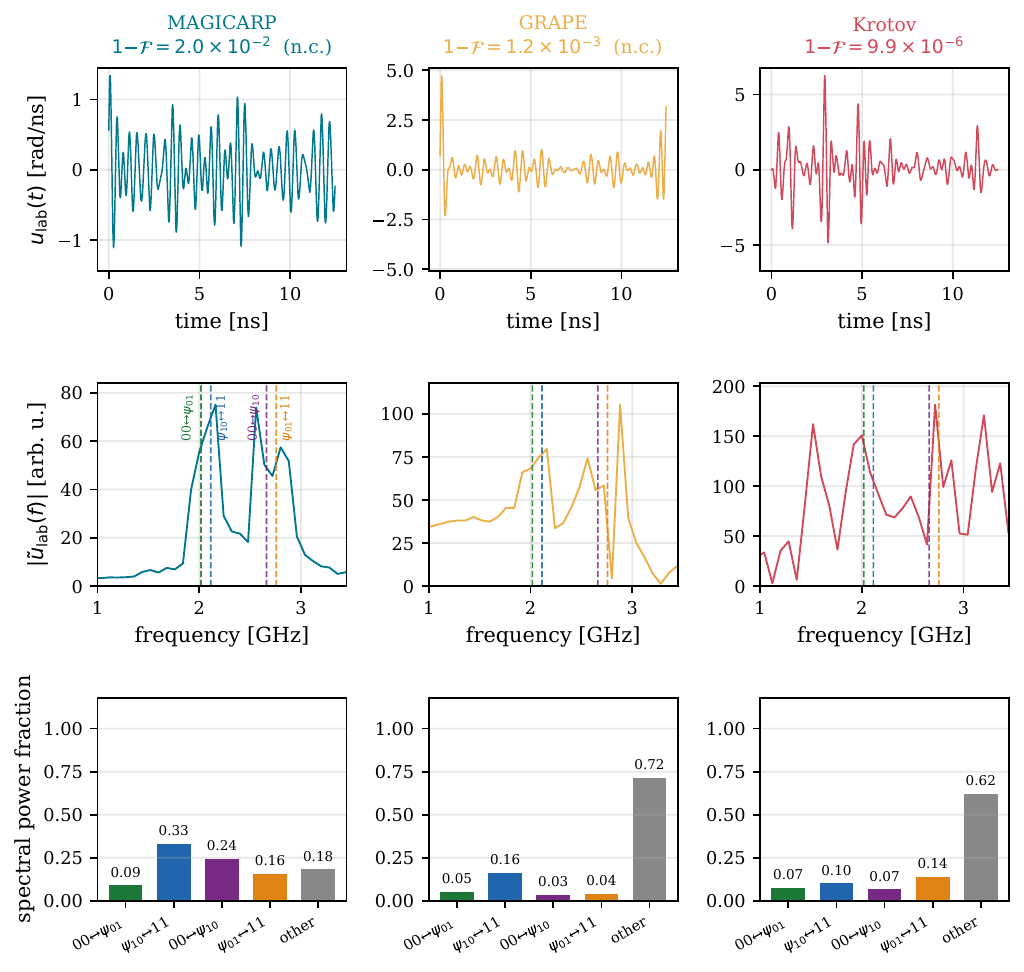}
\caption{The QFT anatomy at $T=12.5\,\mathrm{ns}$, below the weak-driving threshold
(independent vertical scales per panel; at the grid resolution of this short
gate the neighbouring transition peaks merge, so the per-transition windows of
the bottom row abut at the midpoints between transitions). Only Krotov reaches
the $10^{-5}$ halt, with a pathological pulse: $E_2=19.6\,\mathrm{rad^2/ns}$ ($55\times$ the
$T=50\,\mathrm{ns}$ minimum-energy pulse) and $\|u_{\mathrm{lab}}\|_\infty=6.25\,\mathrm{rad/ns}$, comparable
to the dressed transition angular frequencies, with $62\%$ of its
power off every dressed transition. MAGICARP stays in the low-amplitude class
and does not converge ($2.0\times10^{-2}$); GRAPE ends at $1.2\times10^{-3}$,
its pulse already partly degraded toward the broadband regime
($\|u\|_\infty=4.7\,\mathrm{rad/ns}$). The picture is threshold-robust: at a $10^{-3}$ halt
Krotov again converges only at $E_2=16.5\,\mathrm{rad^2/ns}$, $\|u\|_\infty=5.4\,\mathrm{rad/ns}$, and GRAPE again
ends at $1.2\times10^{-3}$ (Section~S3 of the Supporting Information). n.c. indicates the optimization didn't converge to the desired threshold of $10^{-5}$ in the allotted number of iterations.}
\label{fig:qft-spec3x3-short}
\end{figure}

\paragraph{The $T=12.5\,\mathrm{ns}$ anatomy.} At the shortest gate time only Krotov reaches
$10^{-5}$ --- and the price is instructive (\cref{fig:qft-spec3x3-short}):
$E_2 = 19.6\,\mathrm{rad^2/ns}$, i.e.\ $55\times$ the energy of the $T=50\,\mathrm{ns}$ MAGICARP pulse and
$9\times$ the bounded MAGICARP pulse at the same $T$, with peak amplitude
$6.25\,\mathrm{rad/ns}$ --- half the dressed transition angular frequencies
($\Omega_p\approx12.7$--$17.3\,\mathrm{rad/ns}$), far
outside the weak-driving regime. The amplitude-bounded MAGICARP search
saturates at $1-\mathcal{F}\approx2\times10^{-2}$, and GRAPE --- which stays in
the low-energy basin it grew from the zero pulse --- reaches
$1.2\times10^{-3}$ within its $1000$-step budget. The three outcomes together
locate the QFT gate-time threshold precisely: at $T=12.5\,\mathrm{ns}$ a low-amplitude QFT
pulse no longer exists, and the gate can only be bought with an
order-of-magnitude energy premium (the threshold itself is mapped in
\cref{sec:magicarp-sweep}). A two-stage treatment does not change this picture. Warm-started from MAGICARP's stage-one pulse, Krotov stalls at $2.2\times10^{-3}$ while its energy climbs to $8.7\,\mathrm{rad^2/ns}$, and warm-started GRAPE ends at the same $1.3\times10^{-3}$ as its zero-initialized counterpart.

\paragraph{Threshold robustness of the $T=12.5\,\mathrm{ns}$ picture.} This anatomy is not
an artifact of the chosen accuracy. When the halting threshold is loosened by
two decades, from $10^{-5}$ to $10^{-3}$, GRAPE and Krotov behave the same way
at $T=12.5\,\mathrm{ns}$ in terms of infidelity: Krotov converges again, and again only at
enormous cost ($E_2=16.5\,\mathrm{rad^2/ns}$, $\|u\|_\infty=5.4\,\mathrm{rad/ns}$ at the $10^{-3}$ halt vs $19.6$
and $6.25$ at $10^{-5}$), while GRAPE ends at the \emph{same} $1.2\times10^{-3}$
at both halts --- missing the loose target by less than $20\%$ and the strict
one by two decades --- and MAGICARP saturates at $2$--$3\times10^{-2}$ in its
low-amplitude class regardless of the halt. The full $10^{-3}$ study is given
in Section~S3 of the Supporting Information; the method ranking it produces is identical.

\subsection{Robustness to exchange-coupling errors at matched error}
\label{sec:robustness}

\begin{figure}[htbp]
\centering\includegraphics[width=.5\linewidth]{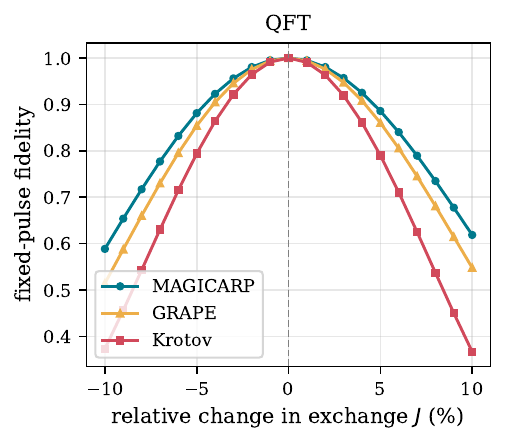}
\caption{Fixed-pulse robustness of the converged $T=50\,\mathrm{ns}$ dressed-QFT pulses to a uniform relative exchange error $\varepsilon \in [-10\%, +10\%]$. The minimal-energy MAGICARP and GRAPE pulses degrade most slowly; the high-energy Krotov solution is the most fragile --- the energy premium buys no robustness.}
\label{fig:fair-robust}
\end{figure}

A pulse is ultimately run on a device whose exchange coupling is known
imperfectly, so the sensitivity of the \emph{fixed} pulse to a $J$ error is an
implementation metric on a par with energy and bandwidth.
\Cref{fig:fair-robust} compares the three converged $T=50\,\mathrm{ns}$ pulses under
a uniform relative exchange error $\varepsilon\in[-10\%,+10\%]$, with the
dressed target rebuilt in the perturbed eigenbasis so that only the pulse, not
the gate definition, is stale (Section~S2 of the Supporting Information).

\paragraph{Low energy is also the most robust.} For the QFT the low-energy MAGICARP and GRAPE pulses are the most tolerant (fidelity $0.881$ and $0.854$ at a $-5\%$ $J$ error, respectively) while the high-energy Krotov solution is the most fragile ($0.794$). Energy bought above the minimum buys no robustness --- the broadband, high-amplitude solution is strictly worse on both axes. The underlying sensitivity is physical rather than method-specific: a
resonant, frequency-encoded pulse of this duration is finely tuned to the
dressed transition frequencies, so a few-percent shift in $J$ detunes the
carriers; robustness can be traded against gate duration or restored by
re-optimizing at the perturbed coupling. We note the boundary of the claim: in
the strong-coupling regime ($JT\sim10^2$) the fixed-pulse
fidelity of \emph{all three} methods collapses within a few percent of exchange
error --- there, robustness is dictated by the accumulated exchange phase, not
by the optimizer (Section~S3 of the Supporting Information).


\section{MAGICARP as a Probe of the quantum speed limit in the weak-amplitude driving regime}
\label{sec:magicarp-sweep}

The comparisons of \cref{sec:exp-advantages} report, at each gate time, the
\emph{single best} pulse returned by a search over amplitude bounds and random
restarts. That best-of-search hides the structure of the underlying control
landscape: how reliably the solver finds a high-fidelity pulse, how the optimal
control energy scales with the available time, and how sharply both change as
the gate is compressed. Because MAGICARP (and the GRAPE it
coincides with) stays in the weak-amplitude pulse class rather than cheating
past the threshold with high-energy pulses (\cref{sec:pathological}), the
raw statistics of its outcomes are a clean instrument for \emph{measuring} the
quantum speed limit in the weak-amplitude driving regime and the minimum-energy law, uncontaminated
by high-energy escapes. 

To expose this structure we ran a dense, sweep of drift-aware MAGICARP on the dressed QFT at the moderate-exchange point. To not impose any bias we show every result as individual datapoint rather then the best-case result in the following.


\paragraph{The no-halt sweep.} We fixed the model and dressed-QFT target of
\cref{sec:spinqb}
($H_0 = -8.5\,Z_1 - 6.5\,Z_2 + 0.15\,(X_1X_2+Y_1Y_2+Z_1Z_2)$, scalar control
$H_{\mathrm c}=X_1+X_2$, $\Delta t = 0.02\,\mathrm{ns}$) and swept the gate time over $100$
values spanning $T\in[5,25]\,\mathrm{ns}$ --- the production window $[12.5,25]\,\mathrm{ns}$ plus a
dense extension reaching deep below the expected threshold. At each $T$ we
launched a $4\times24$ grid of independent optimizations: four amplitude
bounds $\{0.10,0.15,0.22,1.0\}\,\mathrm{rad/ns}$ (the box constraint
$|\theta_a|\le b$ on the generator components, \cref{eq:Gtheta}; the optimizer's only
regularization, not a cap on $u_{\mathrm{lab}}$) crossed with $24$ restarts --- a single
deterministic $\theta=0$ baseline start plus $23$ noisy multistarts --- i.e.\
$96$ runs per gate time. Crucially, the verified-fidelity halt of the fair-halting
protocol was \emph{removed}: each run spends its full RWA-multistart and
lab-refinement budget regardless of the fidelity it reaches, and we record the
achieved verified infidelity $1-\mathcal{F}$ and pulse energy
$E_2=\int_0^T |u_{\mathrm{lab}}|^2\,\mathrm dt$ of every run. The sweep thus
samples the raw outcome distribution of the solver, not the envelope of its
best efforts. The completed scan comprises $9600$ runs. 
Throughout this section we call a run \emph{converged} when its verified
infidelity satisfies $1-\mathcal{F}\le10^{-7}$ --- a threshold chosen in the
gap between the machine-precision cloud ($\sim10^{-15}$) and the stalled cloud,
so the classification is insensitive to its exact value.

\begin{figure}[htbp]
\centering\includegraphics[width=\linewidth]{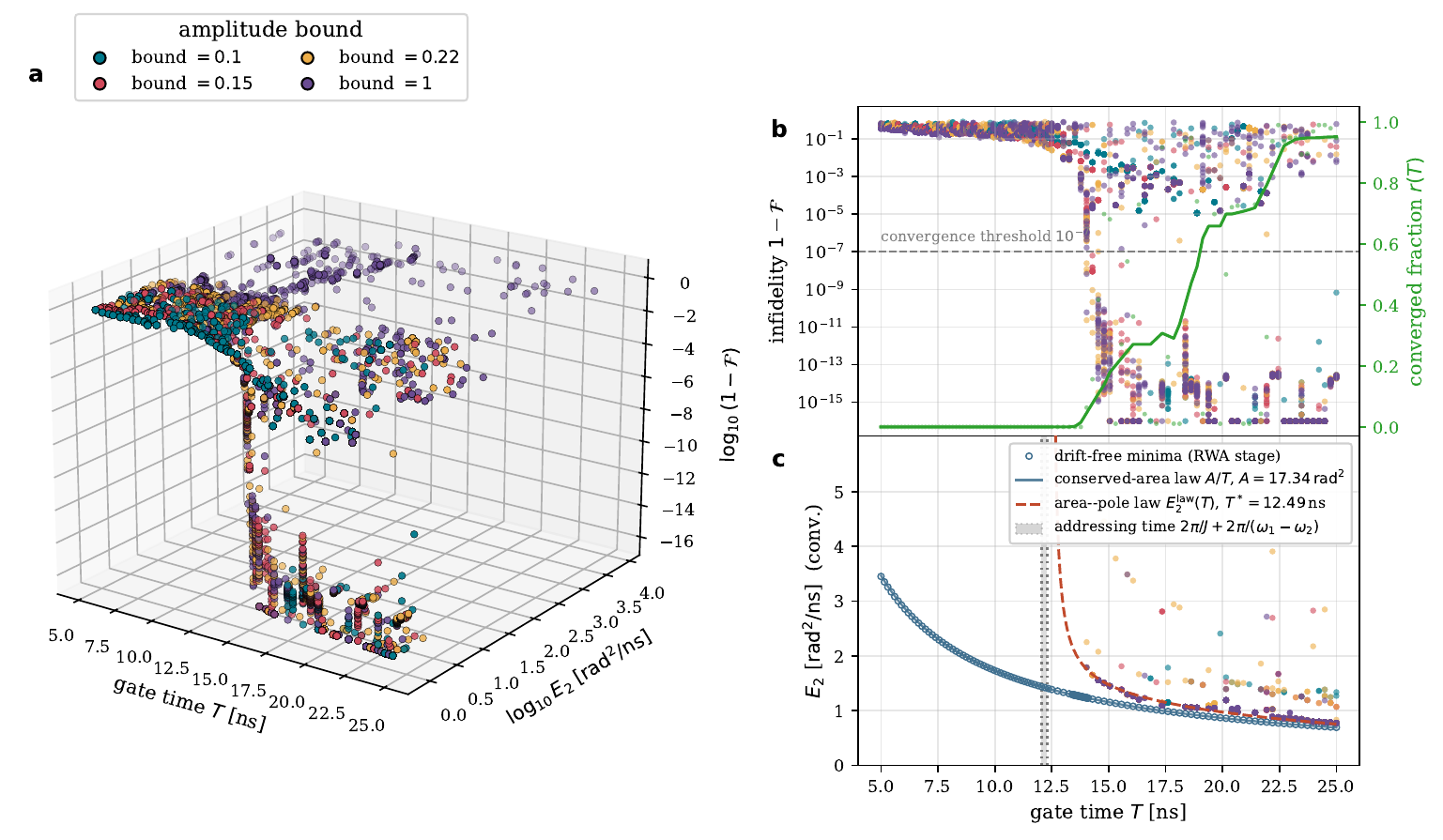}
\caption{The no-halt MAGICARP sweep on the dressed QFT ($9600$ runs: $100$ gate 
times spanning $T\in[5,25]\,\mathrm{ns}$, four amplitude bounds $\times$ $24$ restarts
each; every run kept; colour $=$ amplitude bound, the box constraint $|\theta_a|\le b$ on the
shooting-generator components [\cref{eq:Gtheta}] --- not a cap on the pulse $u_{\mathrm{lab}}$). (a) Joint scatter of $T$,
$\log_{10}E_2$, and $\log_{10}(1-\mathcal{F})$: the converged runs form a tight
low-energy, high-fidelity ridge, sharply separated from the stalled,
high-energy cloud. (b) Infidelity versus $T$ with the $10^{-7}$ convergence
threshold (dashed) and, on the right axis, the fraction $r(T)$ of runs per
gate time below it (points; the line is a running median as a guide to the
eye); no run crosses the threshold below $T\approx14\,\mathrm{ns}$ --- anywhere in the
sub-threshold extension down to $T=5\,\mathrm{ns}$ --- and above it the converged fraction
rises; the points scattered far below the trend (e.g.\ near $T\approx17\,\mathrm{ns}$) are
the basin structure discussed in the text.
(c) $E_2$ versus $T$ for the converged runs only: their lower envelope is the
smooth, bound-independent minimum-energy curve $E_2^{\min}(T)$, rising as
the speed limit is approached; the region $T\lesssim14\,\mathrm{ns}$ is empty of converged runs because no run,
low-amplitude or not, converges there. The dashed curve is the area--pole law
$E_2^{\mathrm{law}}(T)$ of \cref{eq:emin-law}, fitted to the
per-gate-time minimum energies with $A$ fixed at its drift-free value (gate times with too few converged restarts to reach the floor excluded by a one-sided envelope clip, as described in the text). Open markers: the minima of the drift-free problem (dressed RWA stage alone) and their conserved-area law $A/T$, pole-free across the whole window. Dotted vertical line: the addressing time of the shared control, $2\pi/J+2\pi/(\omega_1-\omega_2)\approx12.0$--$12.3\,\mathrm{ns}$ (bare and dressed group splitting); its
parameters and their physical reading are given in the text.}
\label{fig:sweep-overview}
\end{figure}

\paragraph{The solver is stochastic: convergence is a statistic, not a guarantee.}
At essentially every gate time above threshold the runs split into two
well-separated populations (\cref{fig:sweep-overview}a,b): a
\emph{converged} branch that drives the infidelity down to
$1-\mathcal{F}\sim10^{-15}$ (numerical machine precision) and a \emph{stalled}
branch that plateaus near $1-\mathcal{F}\sim10^{-1}$. Which branch a given run
lands in is governed by its starting guess and amplitude bound, not by $T$
alone: as is generic for a shooting method on a non-convex landscape, the
solver descends into one of several basins and one cannot predict in advance
which. Although idealized quantum control landscapes are provably free of suboptimal traps only when the system is controllable and the controls are unconstrained~\cite{rabitz2004,russell2017}, both assumptions are violated here by amplitude bounds and finite gate times, and the existence of traps under constraints remains debated~\cite{pechentannor2011}; we therefore rely on a multistart search. Aggregated over the gate times above the threshold,
$54\%$ of runs ($2280$ of $4224$ at $T\ge14\,\mathrm{ns}$) reach the $10^{-7}$ criterion
--- and none below it --- 
and the single $\theta=0$ baseline start is markedly less reliable than the
random restarts
($14.2\%$ versus $24.2\%$ over the full window), 
which both quantifies why a multistart is necessary and confirms that the
deterministic zero start, while a useful reproducible anchor, is not on its own
a robust strategy. The practical reading is direct: the high-fidelity pulses
reported elsewhere in this work are reliably \emph{findable}, but only as the
best of a handful of restarts --- on average $\sim2$ runs are needed to
obtain one converged pulse above the threshold (the converged fraction rises
to $0.99$ at $T=22.4\,\mathrm{ns}$). 


\begin{table}[htbp]
\centering
\begin{threeparttable}
\caption{Per-bound outcome statistics over the $9600$-run sweep ($2400$ runs
per amplitude bound; the window includes the sub-threshold extension, where
no run can converge). Convergence fractions are with respect to verified
$1-\mathcal{F}\le10^{-7}$. The median converged energy is essentially
bound-independent, whereas the maximum energy --- set by the stalled,
high-amplitude runs --- grows by three orders of magnitude as the bound is
loosened. Energies in $\mathrm{rad^2/ns}$; the dimensionless bound constrains
the generator components. }
\label{tab:sweep-bounds}
\begin{tabular}{ccccc}
\toprule
bound & runs & conv.\ ($\le10^{-7}$) & median $E_2|_{\mathrm{conv}}$ & $\max E_2$ \\
\midrule
0.10 & 2400 & $17.8\%$ & 0.86 & 9.5 \\
0.15 & 2400 & $27.5\%$ & 0.99 & 13.3 \\
0.22 & 2400 & $25.2\%$ & 1.06 & 26.2 \\
1.0  & 2400 & $24.6\%$ & 1.02 & $8.2\times10^{3}$ \\
\bottomrule
\end{tabular}
\end{threeparttable}
\end{table}

\paragraph{A bound-independent minimum-energy optimum.} The converged runs are
far more orderly than the stalled ones. Across all bounds and restarts they
collapse onto a single, narrow minimum-energy floor
(\cref{fig:sweep-overview}c): the converged median energy is
$E_2\approx0.9$--$1.1$ for every amplitude bound (\cref{tab:sweep-bounds}),
and the converged energies over the whole window span
$[0.78,\,4.46]\,\mathrm{rad^2/ns}$ with median $0.98$, the upper tail sitting
at the threshold edge. 
That the floor does not move when the amplitude limit is changed by a factor of
ten --- and, from the fair-halting comparison, that the bound-free,
GRAPE rediscovers the same energy (\cref{sec:spectral})
--- identifies this branch as the genuine minimum-energy solution of the
control landscape rather than an artifact of the dressed ansatz or of any
particular constraint. Selecting the lowest-energy converged run at each $T$
traces a clean, monotonic minimum-energy--versus--time curve
(\cref{tab:sweep-emin}; the lower envelope in
\cref{fig:sweep-overview}c):
$E_2^{\min}$ falls from $1.79$ near the first-convergence gate time ($T\approx14.0\,\mathrm{ns}$) to $0.78\,\mathrm{rad^2/ns}$ at
$T=25\,\mathrm{ns}$, the latter matching the $T=25\,\mathrm{ns}$ fair-halting MAGICARP/GRAPE energies of
\cref{tab:qft-fair} ($0.776$/$0.753$). Out of a stochastic,
basin-hopping solver, then, a deterministic physical law emerges once one
conditions on success: the longer it takes to reach the gate, the gentler and lower-energy the
optimal pulse, because the always-on exchange $J$ has more time to mediate the
entangling rotations of the QFT at weak drive.


\begin{table}[htbp]
\centering
\begin{threeparttable}
\caption{Minimum converged energy $E_2^{\min}(T)$ (lowest-energy run reaching
$1-\mathcal{F}\le10^{-7}$ at each gate time).   The curve is smooth and monotonic and rises steeply on approach to the first-convergence gate time near $T\approx14.0\,\mathrm{ns}$. $E_2^{\min}$ is in $\mathrm{rad^2/ns}$} 
\label{tab:sweep-emin}
\begin{tabular}{cccccccc}
\toprule
$T$ (ns) & 25.0 & 22.4 & 20.4 & 18.4 & 16.3 & 15.1 & 14.0 \\
\midrule
$E_2^{\min}$ & 0.78 & 0.88 & 0.99 & 1.07 & 1.21 & 1.45 & 1.79 \\
\bottomrule
\end{tabular}

\end{threeparttable}
\end{table}

\paragraph{The area--pole law.} The shape of $E_2^{\min}(T)$ is itself
physically transparent, and we condense it into a two-term two-pole model --- the
\emph{area--pole law} ---
\begin{equation}
E_2^{\mathrm{law}}(T)\;=\;\frac{A}{T}\;+\;\frac{B}{T-T^*},
\label{eq:emin-law}
\end{equation}
whose constants have important meanings. The first term is an \emph{area law}.
In the weak-driving regime the target fixes the generalized Rabi angles the
pulse must execute on the dressed transitions, so the minimum pulse
\emph{area} $E_1=\int_0^T|u_{\mathrm{lab}}|\,dt$ is conserved --- the same
$E_1\approx3.5$--$4\,\mathrm{rad}$ found at every gate time in
\cref{sec:pathological}. Delivering a fixed area in a time $T$ costs the least energy when the pulse amplitude is spread evenly over the gate, giving the bound $E_2 \geq E_1^2/T$ (the time-optimal, or geodesic, limit~\cite{carlini2006}). This fixes the $1/T$ scaling and identifies $A$ with the squared pulse area up to an order-unity shape factor that depends on the carrier structure. The second
term is a \emph{pole at the weak-amplitude speed limit}: as $T\to T^{*+}$ the pulse's spectral width $\sim2\pi/T$ becomes comparable to the exchange splitting $J$ of the dressed-transition pairs it must drive differently (next paragraphs), and the crosstalk this produces has to be compensated by pulse shaping (cf.\ Ref.~\cite{motzoi2009}). If that compensation is an additional control path of length $\ell$ that can only be executed in the margin $T-T^*$ by which $T$ exceeds the addressing limit, the same constant-speed argument gives a cost $\ell^2/(T-T^*)$, hence $B=\ell^2$; this is the form the inequality suggests rather than a derivation (we have no reduced model for $B$), and the law is semi-empirical: $A$ measured independently, $T^*$ located at an identified scale, $B$ fitted. We measure $A$ on the drift-free problem --- the dressed RWA stage of \cref{sec:workflow} run alone, in which the dressed quadratures act as independent controls and no explicit drift or carrier appears, its energy being that of the laboratory pulse reconstructed through \cref{eq:lab_pulse_dressed} --- which converges at every gate time down to $T=5\,\mathrm{ns}$ and obeys the pure conserved-area law $E_2\,T=A$ with $A=17.34\pm0.01\,\mathrm{rad^2}$ over $136$ gate times ($R^2>0.9999$), with no pole anywhere in $T\in[5,25]\,\mathrm{ns}$ and the same value at every convergence criterion from $10^{-3}$ to $10^{-7}$. With $A$ fixed, the two remaining parameters fitted to the drifted minima at the $10^{-7}$ criterion --- gate times whose minimum lies more than $1.5\sigma$ above the running envelope (too few converged restarts to reach the floor) iteratively excluded, a minimum envelope having no downward outliers; uncertainties from a bootstrap over the pool of restarts --- are $B=0.81\pm0.04\,\mathrm{rad^2}$ and $T^*=12.49\pm0.09\,\mathrm{ns}$ ($\Delta\chi^2\le1$ interval $12.40$--$12.56\,\mathrm{ns}$); the same procedure on the independent amplitude-unbounded GRAPE envelope gives $B=0.55\pm0.01\,\mathrm{rad^2}$ and $T^*=13.07\pm0.01\,\mathrm{ns}$ ($13.10\,\mathrm{ns}$ with $A$ free). Three consistency checks anchor this reading: the drift-free
$A$ lies just below the directly measured plateau $E_2^{\min}\,T = 20.1\pm1.0\,\mathrm{rad^2}$ that
holds for every inlier gate time $T\ge16\,\mathrm{ns}$ of the sweep, the difference being the tail of the pole term, $BT/(T-T^*)\approx1.6$--$3.7\,\mathrm{rad^2}$ between $25$ and $16\,\mathrm{ns}$; 
$\sqrt{A}=4.2\,\mathrm{rad}$ matches the conserved area $E_1$ up to the multi-tone form
factor; and the pole lies within a nanosecond above the addressing time of the shared control, $2\pi/J+2\pi/(\omega_1-\omega_2)\approx12.0$--$12.3\,\mathrm{ns}$, and $0.2$--$1.4\,\mathrm{ns}$ below the convergence cliff located independently by the outcome statistics ($13.3\,\mathrm{ns}$ for GRAPE, $13.9\,\mathrm{ns}$ for the bounded shooting solver). The residual length
$\sqrt{B}\approx0.7$--$0.9\,\mathrm{rad}$ is a modest compensation path, a fraction of the conserved area $E_1$; we caution that $B$ and $T^*$ are
correlated within the fit window, so the pole's strength should be read as an order of magnitude, and its location --- whose statistical errors are far smaller than the $0.6\,\mathrm{ns}$ spread between solvers and criteria --- to $\approx\pm0.3\,\mathrm{ns}$; we therefore quote $T^*=12.5$--$13.1\,\mathrm{ns}$ throughout. Also, this law is consistent with results from~\cite{tinoco:hal-05404999} which described how increasing the pulses' energies beyond a certain threshold (That they identified as a ratio of the drive to the drift) is not useful to shorten pulses' length.

\paragraph{A low-energy speed-limit threshold.} Read along the time axis, the
sweep resolves the gate-time threshold of \cref{sec:pathological} into a
sharp onset (\cref{fig:sweep-overview}b). Across the entire sub-threshold
extension of the scan --- $5\le T\lesssim13.9\,\mathrm{ns}$, more than five thousand
independent optimizations --- \emph{no} run reaches the convergence
criterion, and for $T\lesssim13.8\,\mathrm{ns}$ none reaches even
$1-\mathcal{F}\le10^{-3}$ in the sweep; the best fidelity attainable at $T=12.5\,\mathrm{ns}$ is
$\mathcal{F}=0.977$ ($1-\mathcal{F}=2.3\times10^{-2}$), consistent with the
bounded-amplitude restart search of \cref{sec:pathological}. The first run of the sweep to
cross $10^{-3}$ appears at $T=13.78\,\mathrm{ns}$, the first to satisfy the $10^{-7}$
criterion at $T=14.03\,\mathrm{ns}$; the converged fraction $r(T)$ exceeds $1/2$ by
$T=14.8\,\mathrm{ns}$ ($67/96$) and reaches $0.99$ at $T=22.4\,\mathrm{ns}$.  
Here $r(T)$ counts only the weak-driving runs, i.e.\ those outside the stalled high-amplitude basin discussed in the next paragraphs ($E_2\le50\,\mathrm{rad^2/ns}$ and peak amplitude $\le10\,\mathrm{rad/ns}$). This exclusion removes no converged run. The fraction rises from $0/94$ at $T=13.78\,\mathrm{ns}$ to $6/95$ at $14.03\,\mathrm{ns}$, $15/96$ at $14.29\,\mathrm{ns}$ and $40/96$ at $14.54\,\mathrm{ns}$. The onset is robust to everything we varied. We extended the sweep to $96$ restarts per bound at all $100$ gate times ($38\,400$ runs in total, including the original $24$ restarts) and added a fine scan in $0.02\,\mathrm{ns}$ steps across $13.4$--$14.1\,\mathrm{ns}$ ($24$ restarts $\times$ five bounds, the fifth bound being $0.5$). Of these $22\,300$ additional runs, none reaches $1-\mathcal{F}\le10^{-3}$ below $T=13.6\,\mathrm{ns}$; the best result at $T=12.5\,\mathrm{ns}$ improves only to $\mathcal{F}=0.982$. The first crossings move slightly, to $13.60\,\mathrm{ns}$ ($10^{-3}$), $13.90\,\mathrm{ns}$ ($10^{-5}$), $13.92\,\mathrm{ns}$ ($10^{-7}$), $13.94\,\mathrm{ns}$ ($10^{-9}$) and $13.96\,\mathrm{ns}$ ($10^{-12}$): tightening the criterion by nine decades shifts the onset by only $0.36\,\mathrm{ns}$. The onset is not an artefact of the shooting parametrization either. As an independent check, we ran zero-initialized GRAPE in a piecewise-constant (pixel) basis on the same $100$ gate times, with $12$ restarts plus a $20$-point scan across the onset ($1344$ runs), using $1000$ L-BFGS-B iterations, no halting, no amplitude bound and no rotating-wave approximation at any stage. It finds nothing below $12.5\,\mathrm{ns}$, even with peak amplitudes up to $26\,\mathrm{rad/ns}$; it first crosses $10^{-3}$ at $12.76\,\mathrm{ns}$ and first converges at $13.27\,\mathrm{ns}$, and above threshold it reproduces the sweep's minimum-energy envelope. The two onsets differ by $0.65\,\mathrm{ns}$ ($0.85\,\mathrm{ns}$ at the $10^{-3}$ criterion), the shooting solver's being the later; where both converge they find the same pulse, so the bounded shooting solver's onset is an upper bound on the physical threshold, tight to $\approx0.7\,\mathrm{ns}$ for the QFT.
This onset has the character of a quantum speed
limit~\cite{caneva2009,deffner2017} --- a minimum duration below which the
target cannot be synthesized --- but it is a \emph{weak-driving} speed limit
rather than an absolute controllability bound.  In contrast to the textbook unitary speed limits, which are fixed by the energy spread or the mean energy above the ground state~\cite{mandelstamtamm1945,margoluslevitin1998}, the limit probed here is a driving (control) speed limit set by the addressing time of the single shared control. An unconstrained Krotov
optimization does reach $1-\mathcal{F}\le10^{-5}$ at $T=12.5\,\mathrm{ns}$, but only by
driving an order of magnitude harder ($E_2\approx19.6\,\mathrm{rad^2/ns}$, peak amplitude
$\approx6.2\,\mathrm{rad/ns}$, comparable to the dressed transition angular frequencies;
\cref{sec:pathological}, \cref{fig:qft-spec3x3-short}).

Equivalently, in the conjugate picture of minimal gate time versus control
strength, a drift-set minimal time that persists at arbitrarily strong
control~\cite{tinoco2025} corresponds precisely to a divergence of the
required energy as $T$ approaches it from above; the pole fitted here is
this divergence, restricted to the weak-amplitude class (the absolute bound
lies lower, as the high-amplitude Krotov point of
\cref{sec:pathological} shows). Exact full-control gate speed limits under a spectral-width budget~\cite{nelson2026} bound the opposite, unconstrained corner of the problem; with a fixed uncontrollable drift and a single control line, exact minimal times are generally out of reach of closed forms, precisely the regime the sweep measures. The measurement is moreover constructive: it returns the minimal smooth pulses that operate at the edge, which a time bound alone cannot provide.

\paragraph{The threshold is the addressing time of the shared control.} The location of the threshold is set by the drift, through the time the single shared control needs to address the dressed transitions separately.
The sub-threshold window $8\lesssim T\lesssim12.3\,\mathrm{ns}$ lies entirely below the addressing time introduced below --- $96$ restarts per gate time find nothing there in the sweep, nor do $384$ in the extended campaign --- and convergence switches on at $T=13.9$--$14.0\,\mathrm{ns}$ (previous paragraph). What amplitude cannot buy is selectivity, because the four dressed transitions are \emph{spectrally crowded}: they sit closer together than the bandwidth of the pulse that has to tell them apart. The two transitions of each qubit, each conditional on the state of the other qubit, are split by exactly $J=0.60\,\mathrm{rad/ns}$ and the two qubit groups by $\omega_1-\omega_2$; a pulse of duration $T$ has a spectral width $\sim2\pi/T$, so it can drive the two members of a $J$-split pair differently only if $T\gtrsim2\pi/J=10.47\,\mathrm{ns}$, and, if the qubit-selective and transition-selective steps cannot be interleaved, resolving the two qubit groups adds $2\pi/(\omega_1-\omega_2)=1.57\,\mathrm{ns}$ ($1.83\,\mathrm{ns}$ with the smallest gap between the groups, $16.72-13.28=3.44\,\mathrm{rad/ns}$)~\cite{vandersypen2005,freeman1998}: the addressing time of the shared control is $2\pi/J+2\pi/(\omega_1-\omega_2)\approx12.0$--$12.3\,\mathrm{ns}$ ($2\pi/J=10.5\,\mathrm{ns}$ with full interleaving). The prefactor of such a time--bandwidth criterion is conventional, and the estimate resolves the crowding only to leading order, so this scale should be read to $\approx\pm20\%$; what the data fix is that the fitted poles sit at $\approx2.4$--$2.5\,\pi/J$. The limit is one of addressability, not of the dressed stage's own approximation: the converged pulses have peak amplitudes of $0.4$--$2\,\mathrm{rad/ns}$ and no weak-driving solution of this work exceeds $\approx6\,\mathrm{rad/ns}$, against dressed transition frequencies of $12.7$--$17.3\,\mathrm{rad/ns}$, so the single-tone rotating-wave approximation holds on both sides of the threshold; and the threshold itself is not an artefact of the dressed stage, since amplitude-unbounded GRAPE in the laboratory frame, which makes no rotating-wave assumption at all, places it only $0.7\,\mathrm{ns}$ lower.
The amplitude ceiling, finally, does not create the
threshold: the onset is the same for bounds $0.15$, $0.22$, and the
effectively unconstrained $1.0$ (first convergence at $T=14.0$--$14.3\,\mathrm{ns}$ in the sweep, $13.92$--$14.00\,\mathrm{ns}$ with $96$ restarts per bound, and $13.94\,\mathrm{ns}$ for a fifth bound $0.5$ in the fine scan);
only the tightest bound $0.10$ delays it (to $T=16.8\,\mathrm{ns}$ in the sweep, $15.8\,\mathrm{ns}$ with $96$ restarts), consistent with the
near-threshold minimal pulse requiring more amplitude than that cap
allows. 
The threshold mapped here is therefore the boundary of the \emph{low-energy}
solution class: below the onset $T\approx13.3$--$13.9\,\mathrm{ns}$ the minimum-energy QFT pulse
ceases to exist, its energy diverging towards the pole $T^*\approx12.5$--$13.1\,\mathrm{ns}$ of \cref{eq:emin-law}, and the gate survives only at an order-of-magnitude energy
premium. Its value lies within a nanosecond above the addressing time of the shared control, $2\pi/J+2\pi/(\omega_1-\omega_2)\approx12.0$--$12.3\,\mathrm{ns}$: the pole $B/(T-T^*)$ is the energy cost of compensating, by pulse shaping, the inter-tone crosstalk that appears once $2\pi/T$ is comparable to $J$, and the gap between pole and first converged run is an amplitude effect --- $0.2\,\mathrm{ns}$ for unbounded GRAPE, $1.4\,\mathrm{ns}$ for the bounded shooting solver, whose cap cannot pay the diverging compensation cost in the last nanosecond above the pole. We note three caveats. First, the onset is not perfectly monotonic in $T$ with $24$ restarts: $T=13.01$ and $T=13.27\,\mathrm{ns}$ reach $\sim7$--$8\times10^{-3}$ while the intervening $T=13.52\,\mathrm{ns}$ falls back to $5\times10^{-2}$ (with $96$ restarts and the fine scan the sub-threshold best falls monotonically, $6.9$, $6.8$ and $2.6\times10^{-3}$ at $13.01$, $13.27$ and $13.52\,\mathrm{ns}$), and a few gate times above the onset retain converged fractions $\le2/96$ (e.g.\ $T\approx17.1$ and $21.2\,\mathrm{ns}$) --- the same basin scatter that produces the two populations. Second, the propagation grid is held at $\Delta t=0.02\,\mathrm{ns}$ throughout;
re-optimization at finer grids shows the short-$T$ residual is only weakly
grid-dependent (Section~S3 of the Supporting Information, grid behaviour), so the
threshold is dominated by the gate-time/amplitude physics rather than by
discretization. Third, the two-pair addressing picture is specific to the detuned regime $\omega_1-\omega_2\gg J$: at strong exchange (the surface-qubit regime of the Supporting Information) the exchange splitting is no longer the smallest scale and the addressing time shrinks with it; the addressing reading has been tested at a single operating point, and its scaling with $J$ is a prediction.

\paragraph{The amplitude bound caps the damage, it does not buy the solution.}
A central, and initially counter-intuitive, message of the sweep concerns
the role of the amplitude bound. The bound has almost no effect on whether the
solver succeeds: the converged fraction is
$18$--$28\%$ across the four bounds, with the tightest bound ($0.10$) the
\emph{worst} and the intermediate $0.15$ the best
(\cref{tab:sweep-bounds}). 
What the bound controls is the \emph{failure mode}. The stalled,
high-amplitude basin that produces the upper population is the runaway
behaviour of the lab-frame refinement; a tight bound truncates it (maximum
energy $E_2=9.5\,\mathrm{rad^2/ns}$ at bound $0.10$), whereas the very loose bound $1.0$ lets it
run away to $E_2\approx8.2\times10^{3}\,\mathrm{rad^2/ns}$, nearly four orders of magnitude
above the minimum-energy floor. The converged solutions at bound $1.0$
nonetheless still sit on the same $E_2\approx1\,\mathrm{rad^2/ns}$ floor as the tightly bounded
ones: loosening the constraint widens the basin of catastrophic failure
without improving the rate of success. The practical recommendation that
follows is exactly the production choice of a tight bound ($\sim0.1$--$0.2$):
it costs nothing in attainable fidelity or optimal energy, while bounding the
energy of the runs that miss and so making the best-of-restarts selection
cheap and safe. \Cref{fig:sweep-overview}a makes the whole picture
legible at once: failure and energy are correlated:high infidelity entails
high, wasted energy; so that the cost of a missed run is paid twice, in
fidelity and in control effort, while the converged ridge is uniformly low on
both axes.
\section{Synthesis}
\label{sec:synthesis}

Taken together, \cref{sec:exp-advantages,sec:magicarp-sweep}
support one statement made three ways. At
matched verified error, the dressed-carrier construction we use in this proposed drift-aware implementation of MAGICARP reaches the target
with the minimal pulse: minimal energy and conserved area
(\cref{sec:pathological}), spectral power concentrated on the dynamically
useful transitions and nowhere else (\cref{sec:spectral}), and the best
fixed-pulse tolerance to exchange error (\cref{sec:robustness}). The
coincidence of the GRAPE with the MAGICARP pulse on every one
of these axes shows that this minimal pulse is a property of the control
landscape, not of the ansatz --- MAGICARP's contribution is to land on it by
construction, with structurally zero off-transition emission, while Krotov's
converged solutions document what the same fidelity costs when the optimizer
is free to leave the weak-driving class. The division of roles is thus: the energy-optimal solution belongs to the Hamiltonian control problem; the ansatz selects the weak-driving class in which it is searched, and it is that class whose boundary \cref{sec:magicarp-sweep} measures. The solver itself is stochastic:
success is a statistic, multiple initial guesses are mandatory, and the amplitude bound
should be read as a safety cap on the failure mode rather than as a knob that
buys convergence. What the parametrization allows for is the energy-extremal structure and spectral parsimony of the converged pulse, not a better-behaved landscape: we claim no advantage in convergence stability.

The same discipline that makes the pulses implementable makes the method a
measurement instrument. Because no run is allowed to cheat past the first
quantum speed limit with a pathological pulse, the unselected sweep of
\cref{sec:magicarp-sweep} cleanly resolves the boundary of the
weak-amplitude solution class --- no low-amplitude QFT from the bounded solver below the first-convergence gate time $\approx14.0\,\mathrm{ns}$ at
this exchange, an upper bound on a physical threshold that amplitude-unbounded GRAPE places $0.7\,\mathrm{ns}$ lower and that coincides with a characteristic addressing time of the shared control, $2\pi/J+2\pi/(\omega_1-\omega_2)\approx12.0$--$12.3\,\mathrm{ns}$. When conditioned on success, a smooth, bound-independent minimum-energy curve described by the
area--pole law, \cref{eq:emin-law}, rising as the limit is approached, its area term measured independently on the drift-free problem and its pole at $T^*\approx12.5$--$13.1\,\mathrm{ns}$. The
method ranking and both of these structures are robust to the choice of
halting accuracy between $10^{-5}$ and $10^{-3}$
(Section~S3 of the Supporting Information).

Two limitations bound the scope of these conclusions, and chart the natural
next steps. First, the present work is restricted to closed-system dynamics:
relaxation, dephasing, and stochastic noise are not included, although in
spin-qubit platforms hyperfine and charge noise --- and, on surfaces, the
finite spin lifetime --- can be the limiting error
mechanism~\cite{petta2005,surfaceqb2026}, and the scaling of gate errors with
decoherence is platform- and dimension-dependent~\cite{jankovic2024}. A full
benchmark against experiment will require extending the optimization either by
stochastic averaging or by Lindblad propagation~\cite{koch2022}. Second, the
robustness analysis is static (a fixed error in $J$), and the basin
sensitivity of the two-stage refinement is the method's practical weakness:
the combination of a multistart search and a tight amplitude bound manages it but does not remove it.
Combining the drift-aware shooting parametrization with ensemble or risk-sensitive robustness costs is a natural continuation: a concrete route is to sample the uncertain exchange coupling and optimize the average gate fidelity over the ensemble~\cite{likhaneja2006,goerz2014}.

\section{Conclusion}
\label{sec:conclusion}

A drift-aware implementation of the MAGICARP shooting method~\cite{magicarp2025}
has been formulated for closed quantum systems with a fixed internal
Hamiltonian. The control law itself is unchanged; it is embedded in a two-stage workflow built
around the drift: a rotating-wave optimization in the dressed basis of the full
drift Hamiltonian, followed by an exact laboratory-frame refinement in which
the physical pulse is reconstructed from carriers at the dressed transition
frequencies. The approach is suited to platforms where the drift determines
the relevant transition structure, with static exchange-coupled spin qubits
--- semiconductor as well as on surfaces.

We benchmarked the drift-aware pulses under a fair-halting protocol against Krotov's method and GRAPE, on the identical model and verified fidelity. At matched error they realize a $\mathrm{NOT}_2$ and a fully entangling dressed QFT with the minimal pulse: the lowest energy, a conserved pulse area, spectral weight concentrated on the gate-essential dressed transitions (spectator weight down to $10^{-4}$ of the total in the strong-coupling regime), and the slowest degradation (highest robustness) under exchange-coupling errors. GRAPE independently converges, on every one of these
axes, to essentially the same pulse --- identifying the minimal-energy,
spectrally clean solution as a property of the control landscape that the
dressed parametrization reaches by construction --- while Krotov's method
meets the same error criteria at a $5$--$57\times$ energy premium with
broadband, fragile pulses. A $9600$-run unselected sweep on the dressed QFT
then turns the bounded solver into a probe of the quantum speed limit in the weak-amplitude driving regime: low-amplitude solutions cease to exist below
$T\approx13.3$--$13.9\,\mathrm{ns}$, their energy diverging towards a pole at $T^*\approx12.5$--$13.1\,\mathrm{ns}$ that lies within a nanosecond of the addressing time of the single shared control, $2\pi/J+2\pi/(\omega_1-\omega_2)\approx12.0$--$12.3\,\mathrm{ns}$. This holds for both optimizers, and above the threshold the minimum control energy follows
the two-parameter area--pole law of \cref{eq:emin-law}, its area term set by the target's rotation content and its pole by the cost of driving below that resolution.
Extensions to open quantum systems, ensemble robustness, multi-channel control, and higher-dimensional (qudit) registers~\cite{etienney2026,jankovic2024} are natural directions for follow-up work.

\section{Methods}

\threesubsection{Closed-system simulations}\\
All simulations are performed for closed quantum dynamics; relaxation,
dephasing, and stochastic noise are not included. Time evolution is the exact
matrix exponential per step, $U=\prod_k e^{-i(H_0+u_kH_{\mathrm c})\Delta t}$,
on the discrete grids of \cref{sec:spinqb}, in every implementation.

\threesubsection{Implementations and verified fidelity}\\
The drift-aware MAGICARP pipeline is implemented in JAX~\cite{bradbury2018};
the Krotov and GRAPE references use the \texttt{krotov}
package~\cite{goerz2019} and QuTiP's \texttt{pulseoptim}~\cite{johansson2013},
respectively, on the same model, grids, and dressed targets (identical across code bases). All methods are scored by the same verified process fidelity, \cref{eq:hs_fidelity}, evaluated by independent re-propagation of the stored pulse (Section~S2 of the Supporting Information).

\threesubsection{Optimization}\\
The optimization variables are the real coefficients $\bm\theta$ of the
anti-Hermitian generator $G(\bm\theta)$. Both stages minimize their loss with
L-BFGS-B~\cite{byrd1995}; the RWA result initializes the laboratory-frame
refinement; the generator components are bounded (the method's only
regularization), and a small multistart is used throughout. The full
algorithm is given in Section~S1 of the Supporting Information, and the unified
benchmarking protocol (halting rule, restarts, selection rule, observables,
runtime accounting) in Section~S2 of the Supporting Information.

\threesubsection{Spectral analysis}\\
Fourier spectra are computed from the final laboratory frame pulses with a common
sampling, windowing, and normalization convention for all methods.
Gate-relevant dressed-transition bands are defined before any comparison;
peak-integrated power fractions use a \texttt{find\_peaks}/\texttt{peak\_widths}
pass on $|\tilde u|^2$ (SciPy~\cite{virtanen2020}) with per-transition windows
clipped at inter-tone midpoints (Section~S2 of the Supporting Information).

\threesubsection{Reproducibility}\\
The parameters reported in this article comprise the drift Hamiltonian and
coupling model, gate durations and time steps, fidelity definition,
anti-Hermitian basis and transition-retention rule, initialization
distributions, L-BFGS-B tolerances and iteration budgets, restart counts,
amplitude-bound sets, carrier phases and amplitudes
($\alpha_p=2$, $\phi_p=0$), and the Fourier-transform convention. Sweep
statistics are frozen at the completed $9600$-run scan; all sweep-derived
numbers are tagged in the source for regeneration.

\section{Acknowledgments}
The authors acknowledge funding by the Institute for Basic Science under grant IBS-R027-D1. This work of the Interdisciplinary Thematic Institute QMat, as part of the ITI 2021-
2028 program of the University of Strasbourg, CNRS and Inserm, was supported by IdEx
Unistra (ANR-10-IDEX-0002), and by SFRI-STRAT’US project (ANR-20-SIFRI-0012) and
EUR QMAT (QMAT ANR-17-EURE-0024) under the framework of the French Investments
for the Future Program. We acknowledge fruitful discussions with R. Pasquier and K. Lutz.

\clearpage
\bibliographystyle{unsrt}
\bibliography{draft_v2}

\clearpage
\setcounter{section}{0}
\setcounter{subsection}{0}
\setcounter{figure}{0}
\setcounter{table}{0}
\setcounter{equation}{0}
\setcounter{algorithm}{0}
\renewcommand{\thesection}{S\arabic{section}}
\renewcommand{\thesubsection}{\thesection.\arabic{subsection}}
\renewcommand{\thefigure}{S\arabic{figure}}
\renewcommand{\thetable}{S\arabic{table}}
\renewcommand{\theequation}{S\arabic{equation}}
\renewcommand{\thealgorithm}{S\arabic{algorithm}}

\section*{Supporting Information}
\addcontentsline{toc}{section}{Supporting Information}

\noindent This Supporting Information collects the technical material accompanying the main text: the drift-aware optimization algorithm (\cref{sec:app-algo}), the fair-halting benchmark protocol (\cref{sec:methods-fair}), the threshold-robustness study at a looser halt (\cref{sec:app-fair-1e3}), and the single-method production results (\cref{sec:app-solo}). Sections, figures, tables and equations of the Supporting Information carry the prefix S; references to the main text are indicated explicitly.

\newpage

\section{Drift-Aware Optimization Workflow}
\label{sec:app-algo}

Algorithm~\ref{alg:workflow} summarizes the complete two-stage pipeline of
Section~2.3 of the main text. The anti-Hermitian basis $\{B_a\}$ spans the full
$\mathfrak{u}(d)$ algebra unless stated otherwise; both stages use
L-BFGS-B~\cite{byrd1995} with bound constraints on the generator components
(MAGICARP's only regularization --- its role as a cap on the runaway failure
mode of the lab-frame refinement is mapped in
Section~4 of the main text), and the production searches wrap the
pipeline in a small multistart over noisy initializations and amplitude
bounds, retaining the lowest-energy converged run
(\cref{sec:methods-fair}).

\begin{algorithm}[h!]
\caption{Dressed-transition drift-aware MAGICARP workflow}
\label{alg:workflow}
\begin{algolines}
\item Define \(H_{\mathrm{local}}\), \(H_{\mathrm{cpl}}\), \(H_0=H_{\mathrm{local}}+H_{\mathrm{cpl}}\), and \(H_{\mathrm{c}}\)
\item Build the target gate \(U_{\mathrm{gate}}\) and compute \(U_{\mathrm{target}}=U_{\mathrm{gate}}e^{-iH_0T}\)
\item Diagonalize \(H_0\ket{\varepsilon_i}=E_i\ket{\varepsilon_i}\)
\item For each retained dressed transition \(p=(i,j)\): compute \(\Omega_p=E_j-E_i\), build \(A_p=\bra{\varepsilon_j}H_{\mathrm{c}}\ket{\varepsilon_i}\ket{\varepsilon_j}\bra{\varepsilon_i}\), \(H_{x,p}=A_p+A_p^\dagger\), \(H_{y,p}=i(A_p^\dagger-A_p)\)
\item Choose an anti-Hermitian basis \(\{B_a\}\), define \(G(\bm{\theta})=\sum_a\theta_aB_a\), initialize \(\bm{\theta}\)
\item[] \textbf{Stage 1: dressed RWA optimization}
\item \textbf{for} each L-BFGS-B iteration \textbf{do}
\item \algI Set \(U_I=\mathbb{I}\)
\item \algI \textbf{for} each RWA time step \textbf{do}
\item \algII Compute $M=U_I G(\bm{\theta})U_I^\dagger$
\item \algII Compute \(u_{x,p}=\Im\,\mathrm{Tr}[H_{x,p}M]\) and \(u_{y,p}=\Im\,\mathrm{Tr}[H_{y,p}M]\)
\item \algII Build \(H_{\mathrm{RWA}}=\sum_p u_{x,p}H_{x,p}+u_{y,p}H_{y,p}\) and propagate with it \(U_I\)
\item \algI \textbf{end for}
\item \algI Reconstruct \(U_{\mathrm{lab}}^{(\mathrm{approx})}=e^{-iH_0T}U_I(T)\), compute \(\mathcal{L}_{\mathrm{RWA}}\), update \(\bm{\theta}\)
\item \textbf{end for}; store \(\bm{\theta}_{\mathrm{RWA}}\)
\item[] \textbf{Stage 2: full laboratory-frame refinement}
\item Initialize \(\bm{\theta}=\bm{\theta}_{\mathrm{RWA}}\)
\item \textbf{for} each L-BFGS-B iteration \textbf{do}
\item \algI Set \(U_{\mathrm{lab}}=\mathbb{I}\)
\item \algI \textbf{for} each laboratory time step \(t\) \textbf{do}
\item \algII Compute \(U_I(t)=e^{+iH_0t}U_{\mathrm{lab}}(t)\) and \(M=U_I(t)G(\bm{\theta})U_I^\dagger(t)\)
\item \algII Compute \(u_{x,p}(t)=\Im\,\mathrm{Tr}[H_{x,p}M]\) and \(u_{y,p}(t)=\Im\,\mathrm{Tr}[H_{y,p}M]\)
\item \algII Reconstruct \(u_{\mathrm{lab}}(t)=\sum_p\alpha_p[u_{x,p}\cos(\Omega_pt+\phi_p)-u_{y,p}\sin(\Omega_pt+\phi_p)]\)
\item \algII Propagate with \(H_{\mathrm{lab}}(t)=H_0+u_{\mathrm{lab}}(t)H_{\mathrm{c}}\)
\item \algI \textbf{end for}
\item \algI Compute \(\mathcal{L}_{\mathrm{lab}}\), update \(\bm{\theta}\)
\item \textbf{end for}
\item Return \(\bm{\theta}_{\mathrm{lab}}\), \(u_{\mathrm{lab}}(t)\), and \(U_{\mathrm{lab}}(T)\)
\end{algolines}
\end{algorithm}

\section{Fair-Halting Benchmark Protocol}
\label{sec:methods-fair}

Optimal-control methods are commonly compared on the fidelity they reach under
heterogeneous stopping criteria (internal cost thresholds, fixed iteration counts,
or machine-precision convergence), which conflates the quality of the optimum with
the willingness of the optimizer to keep running. We instead benchmark the
drift-aware MAGICARP construction against two standard gradient methods ---
Krotov's method and GRAPE --- under a \emph{unified halting rule}: every
optimization stops the moment its \emph{verified} infidelity crosses
$1-\mathcal{F} \le 10^{-5}$, or when its iteration budget is exhausted. At matched
error the meaningful observables are no longer the fidelities (identical by
construction, up to the per-iteration overshoot) but what each method \emph{spends}
to get there: pulse energy and area, peak amplitude, spectral structure, and
optimizer work. The study covers three operating points: $\mathrm{NOT}_2$ in the
strong-coupling regime~\cite{surfaceqb2026}, and $\mathrm{NOT}_2$
and the dressed QFT at the moderate-exchange point of
Section~2.4 of the main text.

\paragraph{Identical physics across implementations.}
All three optimizers act on the same model: the same drift and control
Hamiltonians, the same dressed target gates $B\,G\,B^\dagger$ built from the drift
eigenbasis, and the same piecewise-constant laboratory grid
($\Delta t = 0.02\,\mathrm{ns}$ for the moderate-exchange cases, $0.01\,\mathrm{ns}$ for the
strong-coupling regime). Time evolution is the exact matrix exponential per step,
$U = \prod_k e^{-i(H_0 + u_k H_{\mathrm c})\Delta t}$, in every implementation
(JAX~\cite{bradbury2018} for MAGICARP; QuTiP-based reference implementations of
Krotov and GRAPE~\cite{goerz2019,johansson2013}), and
the target matrices agree across code bases to the last bit. All methods are scored
by the same global-phase-insensitive process fidelity
$\mathcal{F} = |\mathrm{Tr}(U_{\mathrm{target}}^\dagger U)/d|^2$, evaluated by an
\emph{independent re-propagation} of the stored pulse --- the ``verified''
fidelity --- so no method is graded by its own internal cost function.

\paragraph{Per-method halting.}
GRAPE (QuTiP \texttt{pulseoptim}, L-BFGS-B) is stepped one optimizer iteration at a
time and the verified infidelity is evaluated after every step; Krotov's method
(\texttt{krotov} package, $J_{T,\mathrm{re}}$ functional, $\lambda_a = 0.1$,
flat-top update shape) evaluates it in a per-iteration convergence hook. This is deliberately a black-box configuration (as in Ref.~\cite{surfaceqb2026}); remedies that inject structure --- hybrid pre-optimization~\cite{goerz2015} and spectral/amplitude constraints~\cite{palao2013} --- are not applied here. Both stop
at $10^{-5}$ or after at most $1000$ iterations. MAGICARP is two-staged: the
rotating-frame (RWA) stage runs to its own convergence ($45$ L-BFGS-B iterations
per restart) because its averaged-frame loss is \emph{not} the verified metric ---
early-stopping it on the lab-frame fidelity would both apply the wrong criterion
and degrade the starting point handed to the refinement --- while the
laboratory-frame refinement, whose objective \emph{is} the verified infidelity, is
halted per iteration exactly like GRAPE, within a total budget of $1000$ iterations
split over restart chunks. In practice every converged MAGICARP run halts in the
lab stage, so all three methods stop on the same quantity.

\paragraph{Initialization and restarts.}
Each method runs a four-member multistart: one deterministic baseline plus three
restarts whose initial guess receives a small Gaussian ripple (amplitude $0.01$
under a flat-top window). The baselines are method-natural: a flat-top of amplitude
$u_0 = 0.1$ for Krotov, the zero pulse for GRAPE (any structure in
its converged pulse is therefore discovered, not seeded), and a small fixed
$\theta$ for MAGICARP --- an exactly vanishing $\theta$ is a stationary point of
the $\mathrm{NOT}_2$ RWA objective and would never leave the origin. MAGICARP
additionally scans its native amplitude-bound set on the generator components ---
its only regularization, whose role as a cap on the runaway failure mode of the
lab-frame refinement is mapped in Section~4 of the main text --- whereas Krotov
and GRAPE run unconstrained. This asymmetry is deliberate: each method runs in its
natural best-practice configuration, unconstrained amplitude favors the gradient
methods' raw convergence, and the price they pay for it is precisely the observable
of interest.

\paragraph{Selection rule.}
Among the restarts of a given $(T,\text{method})$ that reach verified
$1-\mathcal{F}\le 10^{-5}$, we keep the one with the \emph{lowest pulse energy}
$E_2 = \int_0^T |u_{\mathrm{lab}}|^2\,dt$; if none converge, we keep (and flag) the
lowest-infidelity run, reporting its best achieved value. The benchmark therefore
asks, per method: \emph{what is the cheapest pulse with which you can reach the
target error?}

\paragraph{Reported observables.}
From the selected pulse we report the verified infidelity, energy $E_2$, area
$E_1 = \int_0^T |u_{\mathrm{lab}}|\,dt$, peak amplitude
$\|u_{\mathrm{lab}}\|_\infty$, and for the $\mathrm{NOT}_2$ cases the
\emph{gate-tone spectral fraction}: the share of squared spectral weight
$|\tilde u(f)|^2$ within $\pm2$ FFT bins of the gate-essential dressed transitions,
versus the same share at the spectator transitions. For the panel figures of
Section~3.1 of the main text we additionally integrate $|\tilde u(f)|^2$ over each
dressed-transition peak with data-driven windows: peak locations and integration
widths are determined by a peak-finding pass on the power spectrum (SciPy
\texttt{find\_peaks}~\cite{virtanen2020} with a relative prominence threshold of
$10^{-3}$, widths at
$0.99$ relative height), each window clipped to the disjoint frequency interval
bounded by the midpoints to the neighbouring transitions, with a fixed
$\pm2$-bin window as fallback (flagged in the figures) when no peak is detected
at a transition; the per-transition share of total spectral power is reported.
Grid convergence is probed by \emph{re-optimizing} at each $\Delta t$ at $T=50\,\mathrm{ns}$
(re-optimization, not resampling: the shooting recursion is grid-dependent, and
for the piecewise-constant methods pulse and propagator share the grid by
construction). Robustness is probed by holding the converged $T=50\,\mathrm{ns}$ pulse fixed,
scaling the exchange part of the drift by $(1+\varepsilon)$,
$\varepsilon \in [-10\%, +10\%]$ ($21$ points), and re-propagating; the dressed
target is rebuilt in the perturbed eigenbasis, so only the pulse, not the gate
definition, is stale.

\paragraph{Runtime accounting.}
Wall-clock comparisons across heterogeneous code bases (JAX versus QuTiP) are
implementation-confounded, so alongside the wall time we count \emph{propagation
work}: one unit equals one full lab-grid gate propagation, with per-method
budgets (Krotov: three per iteration --- forward, backward, verified check; GRAPE:
two per step; MAGICARP: RWA objective evaluations rescaled by the grid ratio, plus
lab-stage evaluations and verified checks). The quotient
$\text{wall time}/\text{prop work}$ then measures optimizer speed per unit of
physics. Under this normalization the per-propagation cost varies within roughly an
order of magnitude across methods and cases (the JAX-jitted shooting is typically
cheapest, Krotov dearest --- up to $\sim\!20\times$ the shooting cost --- while
GRAPE matches the shooting cost in the strongly coupled case), and the gradient
methods need fewer propagations per converged run ($10^2$--$10^3$ versus
$10^3$--$10^4$ for the shooting bound-and-seed search); total wall times to
the halt are comparable, $0.5$--$3$ hours per (case, method) sweep on a single CPU
core. No method is runtime-prohibitive at this system size, and we do not weight
the comparison by speed.

\begin{table}[htbp]
\centering
\caption{Fair-halting protocol summary. All methods stop at verified
$1-\mathcal{F}\le10^{-5}$ (checked per iteration) or at their iteration budget;
among converged restarts the lowest-$E_2$ pulse is kept.}
\label{tab:fair-protocol}
\begin{tabular}{lllll}
\toprule
Method & Baseline guess & Restarts & Halt check & Budget \\
\midrule
MAGICARP & small fixed $\theta$ & $+3$ noisy, $\times$ bound scan & lab stage, per iteration & $45$ (RWA) $/\,1000$ (lab) \\
Krotov & flat-top $u_0=0.1$ & $+3$ noisy & hook, per iteration & $1000$ \\
GRAPE & zero pulse & $+3$ noisy & per L-BFGS-B step & $1000$ \\
\bottomrule
\end{tabular}
\end{table}

\paragraph{The strong-coupling regime at the $10^{-5}$ halt.} For completeness we
record the strict-halt results at the third operating point (parameters as in
Section~2.4 of the main text and \cref{sec:krotovreg-fair-1e3}); the
looser-halt counterpart is \cref{tab:krotovreg-fair-1e3}.

\begin{table}[htbp]
\centering
\caption{$\mathrm{NOT}_2$ in the strong-coupling regime under the fair $10^{-5}$ halt.
$E_2$ in rad$^2$/ns, $\|u_{\mathrm{lab}}\|_\infty$ in rad/ns. Asterisks mark runs that
exhausted their budget without reaching $10^{-5}$ (best value reported). Last column:
share of squared spectral weight within $\pm2$ FFT bins of the two gate-essential dressed
tones (spectator share in parentheses).}
\label{tab:krotovreg-fair}
\begin{tabular}{llccccc}
\toprule
$T$ (ns) & method & $1-\mathcal{F}$ & $E_2$ & $E_1$ & $\|u_{\mathrm{lab}}\|_\infty$ & gate (spec.) frac. \\
\midrule
50   & MAGICARP & $4.5\times10^{-5}\,{}^{*}$ & 0.217 & 2.71 & 0.13 & 0.93 ($8\times10^{-5}$) \\
     & Krotov   & $6.0\times10^{-6}$ & 8.25  & 15.8 & 1.29 & 0.03 (0.01) \\
     & GRAPE    & $5.6\times10^{-6}$ & 0.218 & 2.71 & 0.15 & 0.91 (0.01) \\
\midrule
30   & MAGICARP & $5.8\times10^{-3}\,{}^{*}$ & 2.47 & 7.03 & 0.67 & 0.95 ($2\times10^{-4}$) \\
     & Krotov   & $4.4\times10^{-6}$ & 6.26  & 10.6 & 1.42 & 0.08 (0.23) \\
     & GRAPE    & $9.8\times10^{-6}$ & 0.366 & 2.72 & 0.25 & 0.90 (0.02) \\
\midrule
12.5 & MAGICARP & $4.4\times10^{-4}\,{}^{*}$ & 0.910 & 2.78 & 0.61 & 0.95 (0.03) \\
     & Krotov   & $1.1\times10^{-6}$ & 4.01  & 5.45 & 1.69 & 0.26 (0.55) \\
     & GRAPE    & $9.7\times10^{-6}$ & 0.914 & 2.74 & 0.66 & 0.93 (0.05) \\
\bottomrule
\end{tabular}
\end{table}

\paragraph{Energy and selectivity at the strict halt.} GRAPE converges at every
gate time onto pulses energetically indistinguishable from the MAGICARP ones
($E_2 = 0.218$ vs $0.217$ rad$^2$/ns at $T=50\,\mathrm{ns}$; $0.914$ vs $0.910$
at $12.5\,\mathrm{ns}$), with the same conserved area $E_1\approx2.7$; Krotov
converges everywhere at $4$--$38\times$ the pulse energy. The spectral-fraction
column adds the one observation that distinguishes the strict halt: at
$T=50\,\mathrm{ns}$ the MAGICARP pulse holds $93\%$ of its spectral weight on
the two gate tones and $8\times10^{-5}$ on the spectators, while GRAPE, despite
finding the same energy, leaks $\sim\!10^{-2}$ onto the spectators --- two
orders of magnitude more --- because nothing in its parametrization
distinguishes the tones. The dressed parametrization is therefore not merely
recovering what an unbiased gradient would find: its structural suppression of
spectator drive exceeds the converged GRAPE solution by $\sim\!100\times$ at
equal energy. Under this leaner fair-protocol search the shooting pipeline does
not itself cross $10^{-5}$ at this operating point (it saturates at
$4.5\times10^{-5}$ at $T=50\,\mathrm{ns}$, and its $T=30\,\mathrm{ns}$ point
lands in the high-amplitude basin, $E_2=2.47$ --- a production best-of search
reaches $1.2\times10^{-4}$ with $E_2=0.362$ there); as in the moderate-exchange
case, GRAPE's convergence at MAGICARP's own energy shows the target is
reachable within the low-energy pulse family, and the basin sensitivity of the
lab-frame refinement remains the method's main practical weakness
(Section~4 of the main text).

\section{Threshold Robustness: the Fair-Halting Study at a $10^{-3}$ Halt}
\label{sec:app-fair-1e3}

The comparison of the main text (Section~3) halts every optimizer
at verified $1-\mathcal{F}\le10^{-5}$. To verify that none of its conclusions
is an artifact of the chosen accuracy, the entire study was repeated at a halt
two decades looser, $1-\mathcal{F}\le10^{-3}$, under the otherwise identical
protocol of \cref{sec:methods-fair} (the only further difference: the
grid-convergence study runs at the intermediate gate time $T=25\,\mathrm{ns}$ rather than
$T=50\,\mathrm{ns}$; total wall times to the looser halt drop to minutes--$\sim$2 hours per
case and method). At the looser halt MAGICARP itself converges at most
operating points and GRAPE is as spectrally clean as MAGICARP (spectator
weight $\sim10^{-4}$); the Krotov energy premium ($\approx5$--$57\times$) and
the MAGICARP/GRAPE energy coincidence are unchanged --- the ranking is
threshold-robust.

\subsection{QFT: comparison with Krotov and GRAPE at matched verified error}
\label{sec:qft-fair-1e3}

Under the fair-halting protocol of \cref{sec:methods-fair} with the
halt loosened to $10^{-3}$ --- identical model, grid, target, and verified
metric; every optimizer stopped at verified $1-\mathcal{F}\le10^{-3}$;
lowest-energy converged restart kept --- the dressed QFT separates the three
methods by \emph{control cost}, not by capability
(\cref{tab:qft-fair-1e3}). The same ranking persists when the halt is
tightened by two further decades, so none of the conclusions below is an
artifact of the chosen accuracy.

\begin{table}[htbp]
\centering
\caption{Dressed QFT under the fair $10^{-3}$ halt: verified infidelity, pulse energy
$E_2$, pulse area $E_1$, and peak amplitude for MAGICARP, Krotov, and (zero-initialized)
GRAPE. Asterisks mark runs that exhausted their budget without reaching $10^{-3}$ (best
value reported). Units as in Table~2 of the main text.}
\label{tab:qft-fair-1e3}
\begin{tabular}{llcccc}
\toprule
$T$ (ns) & method & $1-\mathcal{F}$ & $E_2$ & $E_1$ & $\|u_{\mathrm{lab}}\|_\infty$ \\
\midrule
50   & MAGICARP & $9.9\times10^{-4}$ & 0.350 & 3.50 & 0.19 \\
     & Krotov   & $9.2\times10^{-4}$ & 4.82  & 9.23 & 2.08 \\
     & GRAPE    & $6.6\times10^{-4}$ & 0.350 & 3.46 & 0.22 \\
\midrule
25   & MAGICARP & $9.9\times10^{-4}$ & 0.766 & 3.61 & 0.42 \\
     & Krotov   & $9.5\times10^{-4}$ & 6.80  & 8.11 & 2.98 \\
     & GRAPE    & $8.4\times10^{-4}$ & 0.728 & 3.51 & 0.45 \\
\midrule
12.5 & MAGICARP & $2.7\times10^{-2}\,{}^{*}$ & 1.94 & 3.94 & 1.37 \\
     & Krotov   & $9.9\times10^{-4}$ & 16.5  & 9.24 & 5.39 \\
     & GRAPE    & $1.2\times10^{-3}\,{}^{*}$ & 5.61 & 4.92 & 4.73 \\
\bottomrule
\end{tabular}
\end{table}

\paragraph{Two methods find the same pulse; one pays a premium.} At $T=50\,\mathrm{ns}$ and $T=25\,\mathrm{ns}$ all
three optimizers reach the target error, but MAGICARP and GRAPE converge onto essentially
the \emph{same} minimal-energy solution: at $T=50\,\mathrm{ns}$ their energies agree to three digits
($E_2 = 0.350\,\mathrm{rad^2/ns}$ for both) and at $T=25\,\mathrm{ns}$ to $5\%$ ($0.766$ vs $0.728$), with matching areas
and peak amplitudes. The GRAPE, free of any frequency ansatz,
independently rediscovers the four-tone, low-energy pulse that the dressed parametrization
produces by construction --- strong evidence that this solution is the natural
minimal-energy optimum of the control landscape rather than an artifact of the ansatz.
Krotov, climbing from a flat-top guess, satisfies the same error criterion with
$14\times$ ($T=50\,\mathrm{ns}$) and $9\times$ ($T=25\,\mathrm{ns}$) more pulse energy, $\sim\!10\times$ the peak
amplitude, and more than twice the pulse area: at matched fidelity its solution is simply
a more expensive point of the same landscape.

\paragraph{The $T=12.5\,\mathrm{ns}$ threshold.} At the shortest gate time only Krotov reaches
$10^{-3}$ --- and the price is instructive: $E_2 = 16.5\,\mathrm{rad^2/ns}$, i.e.\ $47\times$ the energy of
the $T=50\,\mathrm{ns}$ MAGICARP pulse and $8.5\times$ the bounded MAGICARP pulse at the same $T$,
with peak amplitude $5.4\,\mathrm{rad/ns}$, far outside the weak-driving regime. The amplitude-bounded
MAGICARP search saturates at $1-\mathcal{F}\approx2.7\times10^{-2}$; GRAPE --- which
stays in the low-energy basin it grew from the zero pulse --- ends at
$1.2\times10^{-3}$, missing the halt by less than $20\%$ after its full $1000$-step
budget. The three outcomes together locate the QFT gate-time threshold: at $T=12.5\,\mathrm{ns}$ a
low-amplitude QFT pulse no longer exists, and the gate can only be bought with an
order-of-magnitude energy premium (cf.\ the threshold analysis of
Section~4 of the main text).

\paragraph{Robustness at matched error.} Holding each converged $T=50\,\mathrm{ns}$ pulse fixed and
perturbing the exchange coupling, the low-energy MAGICARP and GRAPE pulses are also the
most tolerant (fidelity $0.889$ and $0.854$ at a $-5\%$ $J$ error, respectively), while
the high-energy Krotov solution is the most fragile ($0.803$). Energy bought above the
minimum buys no robustness.

\paragraph{Grid behaviour.} In the $T=25\,\mathrm{ns}$ re-optimization sweep the piecewise-constant
methods meet the halt at every step size, including the coarsest grid
($\Delta t = 0.1\,\mathrm{ns}$, $250$ steps) --- their pulse lives on the propagation grid by
construction, so any grid is internally consistent --- whereas the carrier-based MAGICARP
reconstruction requires the grid to resolve the $2$--$3\,\mathrm{GHz}$ dressed
carriers: it degrades to $3.0\times10^{-2}$ at $\Delta t=0.1\,\mathrm{ns}$ and converges from
$\Delta t \le 0.05\,\mathrm{ns}$. This is a structural, not a numerical, distinction between the
parametrizations.

\begin{figure}[htbp]
\centering\includegraphics[width=\linewidth]{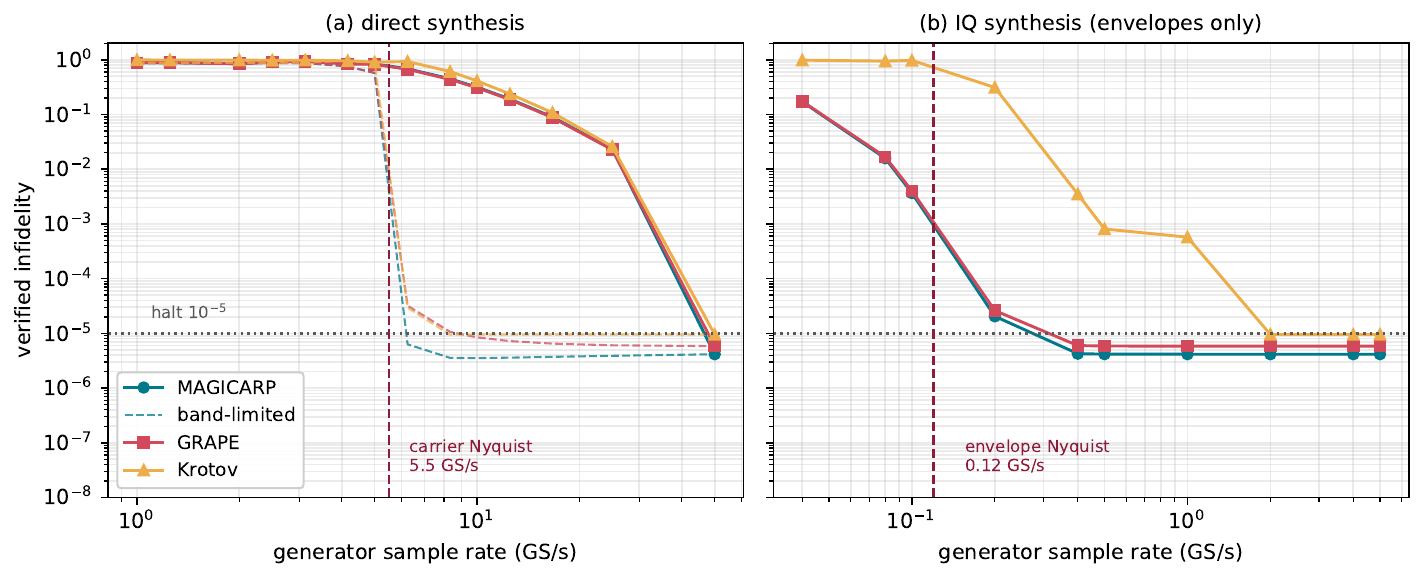}
\caption{Generator temporal resolution for the converged $T=50\,\mathrm{ns}$ dressed-QFT pulses of the three methods at matched verified error (MAGICARP $4.2\times10^{-6}$, GRAPE $5.8\times10^{-6}$, Krotov $9.5\times10^{-6}$). \textbf{(a)} Direct synthesis: the laboratory waveform is handed to the generator as it stands, so the converter must itself resolve the $2$--$3\,\mathrm{GHz}$ oscillation. Solid curves, the raw sample-and-hold staircase with no anti-imaging filter --- the pessimistic bound, and the one under which a pulse must be synthesized as drawn on the optimization grid; dashed curves, only the content below half the sample rate --- the best any generator at that rate could achieve. The latter hold the converged error until the rate falls below twice the highest dressed transition frequency and then fail abruptly, so the binding constraint is the carrier Nyquist condition, identical for all three pulses. \textbf{(b)} IQ synthesis: the generator emits only baseband quadrature envelopes on the four fixed dressed carriers ($2.018$, $2.113$, $2.661$, $2.757\,\mathrm{GHz}$), held piecewise constant and upconverted, so only the envelope bandwidth reaches the converter; the reconstruction is a least-squares projection onto the producible waveform set, applied identically to all three methods. The dressed-carrier pulses hold their converged error down to $0.4\,\mathrm{GS/s}$; the broadband pulse departs from its own at $1\,\mathrm{GS/s}$, a quarter of its norm lying outside the four-carrier span. All reconstructions are re-propagated on the $\Delta t=0.02\,\mathrm{ns}$ grid and scored with the fair-halting verified fidelity. Dotted line, the $10^{-5}$ halting threshold; dashed verticals, the carrier Nyquist rate ($5.5\,\mathrm{GS/s}$) and the envelope Nyquist rate at bandwidth $3/T$ ($0.12\,\mathrm{GS/s}$).}
\label{fig:resolution}
\end{figure}

\paragraph{Generator temporal resolution.} The grid above is the one the optimizer and the propagator share; what an arbitrary-waveform generator must deliver is a separate question, and it separates the parametrizations differently. We take the converged $T=50\,\mathrm{ns}$ dressed-QFT pulses of the three methods at matched verified error ($4.2$, $5.8$ and $9.5\times10^{-6}$ for MAGICARP, GRAPE and Krotov), write each to progressively coarser generator grids along two synthesis routes, and re-propagate the reconstruction on the original $\Delta t=0.02\,\mathrm{ns}$ grid with the verified fidelity of the fair-halting protocol (\cref{fig:resolution}). Synthesized \emph{directly}, the laboratory waveform carries the $2$--$3\,\mathrm{GHz}$ oscillation itself, so the converter must resolve the carrier: with an ideal reconstruction filter all three pulses hold their converged error only down to the carrier Nyquist rate $2\max_p\Omega_p/2\pi=5.5\,\mathrm{GS/s}$, and under plain sample-and-hold all three need the full $50\,\mathrm{GS/s}$ of the optimization grid. On this route a dressed-carrier pulse is no cheaper to draw than a broadband one. Its advantage is that it need not be drawn: Eq.~(10) of the main text delivers it already as fixed carriers at the dressed transition frequencies carrying quadrature envelopes of bandwidth a few times $1/T$, which is the native programming model of IQ-modulated sources and of digital-upconversion electronics --- one envelope per carrier. Asking instead what each pulse costs a generator that emits only those envelopes --- the same sample-and-hold model applied to the baseband channels, the reconstruction obtained by a least-squares projection onto the producible waveform set and applied identically to all three methods --- the dressed-carrier pulses hold their converged error down to $0.4\,\mathrm{GS/s}$, a factor of $14$ below the best direct synthesis and $125$ below the optimization grid, whereas the broadband Krotov pulse leaves its own error already at $1\,\mathrm{GS/s}$, a quarter of its waveform norm lying outside the four-carrier span. The generator burden is therefore a property of the pulse representation rather than of the optimizer, and it is in this sense that the dressed-carrier construction is experimentally cheap.

\begin{figure}[htbp]
\centering\includegraphics[width=\linewidth,trim=0 0 0 50,clip]{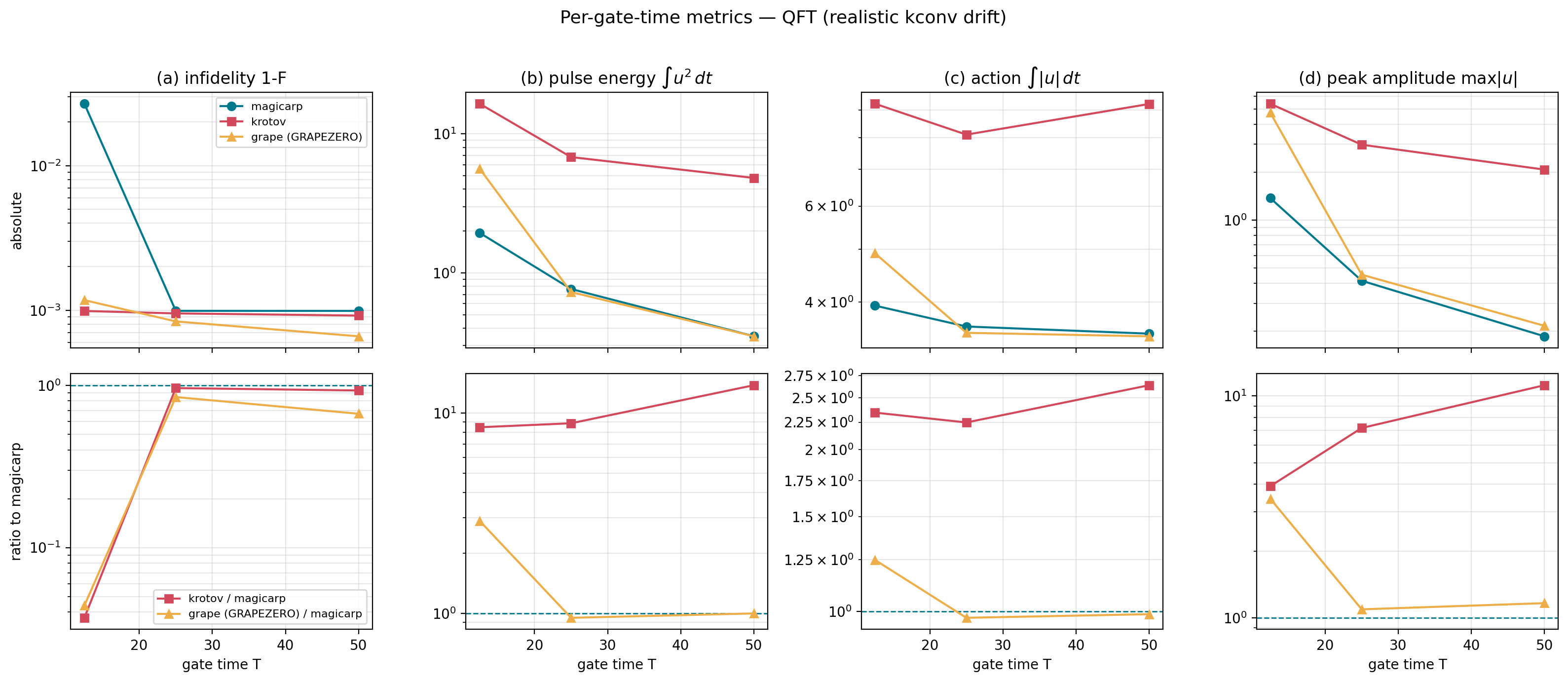}
\caption{Fair-halting comparison for the dressed QFT at the $10^{-3}$ halt: verified
infidelity, pulse energy $E_2$, and peak amplitude versus gate time for MAGICARP, Krotov,
and GRAPE. MAGICARP and GRAPE coincide on the low-energy solution
branch; Krotov converges at every $T$ at an order-of-magnitude energy premium.}
\label{fig:qft-fair1e3-perT}
\end{figure}

\begin{figure}[htbp]
\centering\includegraphics[width=0.62\linewidth]{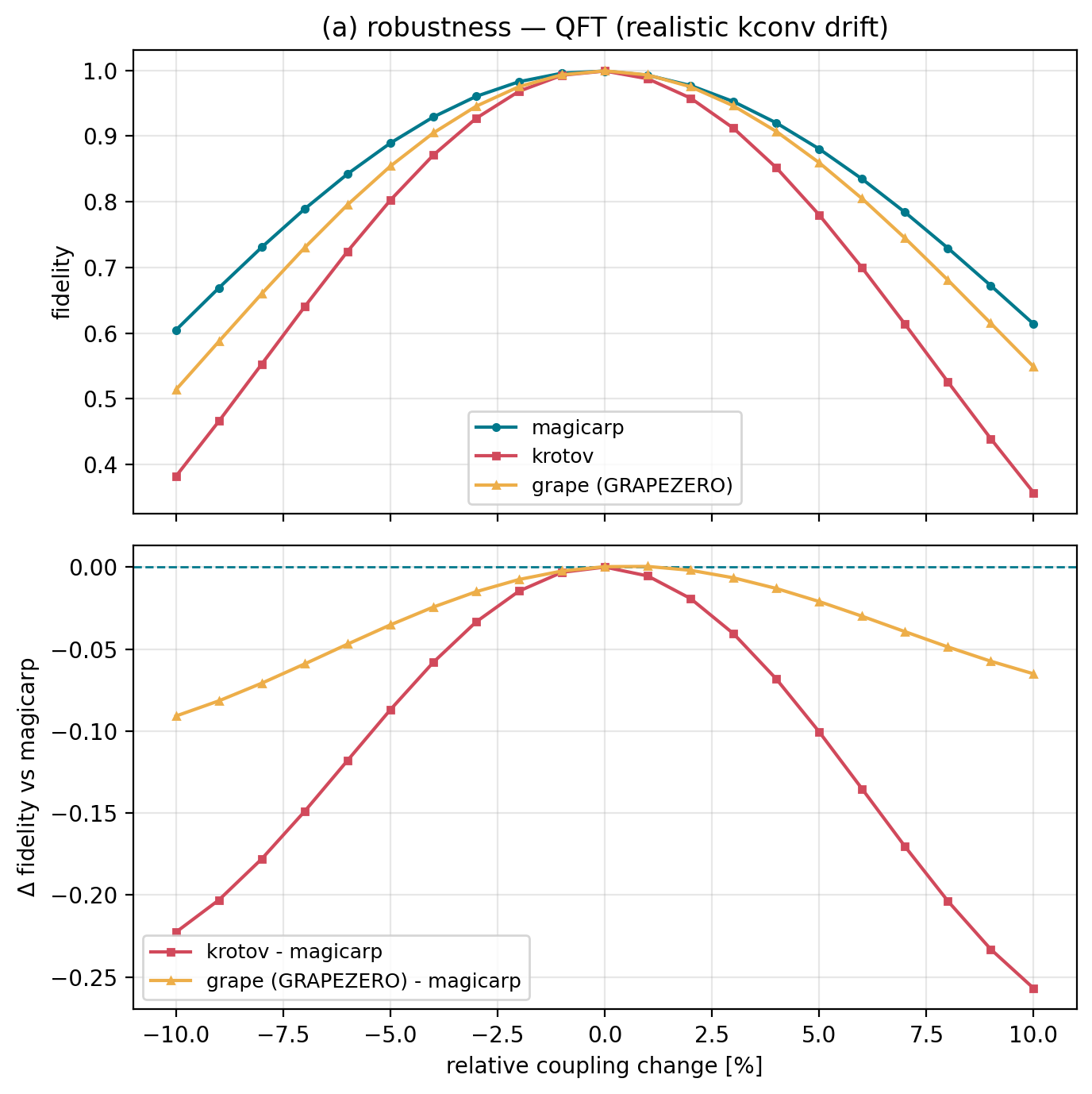}
\caption{Fixed-pulse robustness of the three converged $T=50\,\mathrm{ns}$ QFT pulses to a uniform
exchange error at the $10^{-3}$ halt. The minimal-energy MAGICARP/GRAPE pulses degrade
most slowly; the high-energy Krotov pulse is the most sensitive.}
\label{fig:qft-fair1e3-robust}
\end{figure}

\subsection{$\mathrm{NOT}_2$ in the strong-coupling regime: quantitative head-to-head}
\label{sec:krotovreg-fair-1e3}

The third operating point is the strong-coupling regime of the
$\mathrm{NOT}_2$ benchmark study~\cite{surfaceqb2026}, performed there for closed-system dynamics
(Section~2.4 of the main text), at $\Delta t = 0.01\,\mathrm{ns}$. In this regime the
published benchmark comparison is
qualitative --- our pulse against the reference Krotov spectrum
(\cref{sec:app-solo}). The
fair-halting study makes it quantitative: both reference methods re-implemented
on the identical model, same dressed target, same verified fidelity, all halted
at $10^{-3}$ (\cref{tab:krotovreg-fair-1e3}).

\begin{table}[htbp]
\centering
\caption{$\mathrm{NOT}_2$ in the strong-coupling regime under the fair $10^{-3}$ halt.
$E_2$ in rad$^2$/ns, $\|u_{\mathrm{lab}}\|_\infty$ in rad/ns. The asterisk marks the one
run that exhausted its budget without reaching $10^{-3}$ (best value reported). Last
column: share of squared spectral weight within $\pm2$ FFT bins of the two gate-essential
dressed tones (spectator share in parentheses).}
\label{tab:krotovreg-fair-1e3}
\begin{tabular}{llccccc}
\toprule
$T$ (ns) & method & $1-\mathcal{F}$ & $E_2$ & $E_1$ & $\|u_{\mathrm{lab}}\|_\infty$ & gate (spec.) frac. \\
\midrule
50   & MAGICARP & $9.5\times10^{-4}$ & 0.216 & 2.70 & 0.13 & 0.93 ($3\times10^{-4}$) \\
     & Krotov   & $1.8\times10^{-4}$ & 8.22  & 15.8 & 1.29 & 0.03 (0.01) \\
     & GRAPE    & $9.0\times10^{-4}$ & 0.212 & 2.67 & 0.15 & 0.91 ($1\times10^{-4}$) \\
\midrule
30   & MAGICARP & $5.8\times10^{-3}\,{}^{*}$ & 2.47 & 7.03 & 0.67 & 0.95 ($2\times10^{-4}$) \\
     & Krotov   & $6.1\times10^{-4}$ & 6.20  & 10.6 & 1.42 & 0.08 (0.23) \\
     & GRAPE    & $2.6\times10^{-4}$ & 0.360 & 2.70 & 0.22 & 0.92 ($3\times10^{-4}$) \\
\midrule
12.5 & MAGICARP & $7.7\times10^{-4}$ & 0.866 & 2.70 & 0.53 & 0.99 ($4\times10^{-4}$) \\
     & Krotov   & $1.3\times10^{-4}$ & 4.01  & 5.45 & 1.69 & 0.26 (0.55) \\
     & GRAPE    & $7.0\times10^{-4}$ & 0.862 & 2.70 & 0.53 & 0.98 ($5\times10^{-4}$) \\
\bottomrule
\end{tabular}
\end{table}

\paragraph{Energy at matched error.} The pattern of the moderate-exchange study repeats.
GRAPE converges at every gate time onto pulses energetically indistinguishable from the
MAGICARP ones --- at $T=12.5\,\mathrm{ns}$ the two are virtually the same pulse
($E_2 = 0.862$ vs $0.866$ rad$^2$/ns, $\|u\|_\infty = 0.53$ for both, gate-tone fraction
$0.98$ vs $0.99$), and at $T=50\,\mathrm{ns}$ they agree to $2\%$. Krotov also converges
everywhere --- consistent with the near-unit fidelities of the original benchmark --- but
at $5$--$39\times$ the pulse energy ($4.0$--$8.2$ rad$^2$/ns), up to $5.9\times$ the pulse
area, and $\sim\!10\times$ the peak amplitude. The four-peak, broadband character of the
published Krotov pulse is thus not an artifact of its stopping rule: under a matched
verified halt it persists, and it costs an order of magnitude in energy relative to the
minimal two-tone solution.

\paragraph{Spectral selectivity.} The spectral-fraction column quantifies the parsimony
claim method-by-method. At this halt the MAGICARP and GRAPE pulses are equally clean:
$91$--$99\%$ of their spectral weight sits on the two gate tones with spectator weight at
the $10^{-4}$ level, while Krotov places at most $26\%$ at the gate tones and at
$T=12.5\,\mathrm{ns}$ actually drives the spectators harder than the gate transitions
($0.55$ vs $0.26$). Spectral parsimony at matched error is thus a property of the
\emph{solution family}, discovered by GRAPE and built in by MAGICARP --- and missed
entirely by the flat-top-seeded Krotov flow.

\paragraph{The $T=30\,\mathrm{ns}$ outlier} The one non-converged entry is
MAGICARP at $T=30\,\mathrm{ns}$ ($5.8\times10^{-3}$, $E_2=2.47$): under the leaner
fair-protocol search (one per-$T$ bound, four seeds) the lab-frame refinement lands in
the high-amplitude basin --- the same runaway failure mode mapped by the sweep of
Section~4 of the main text; an unconstrained best-of production search instead reaches
$1.2\times10^{-4}$ with $E_2=0.362$ at this gate time. GRAPE's convergence at
$E_2 = 0.360$ confirms the low-energy solution exists at this gate time; the miss is a
basin-selection failure of the refinement stage, not a property of the pulse family.

\paragraph{Robustness.} In this strongly coupled regime ($J = 5\,\mathrm{rad/ns}$,
$JT\sim10^2$) the fixed-pulse fidelity of \emph{all three} methods collapses within a few
percent of exchange error --- robustness here is dictated by the accumulated exchange
phase, not by the optimizer, and no method's solution is meaningfully more tolerant than
the others'.

\begin{figure}[htbp]
\centering\includegraphics[width=\linewidth,trim=0 0 0 50,clip]{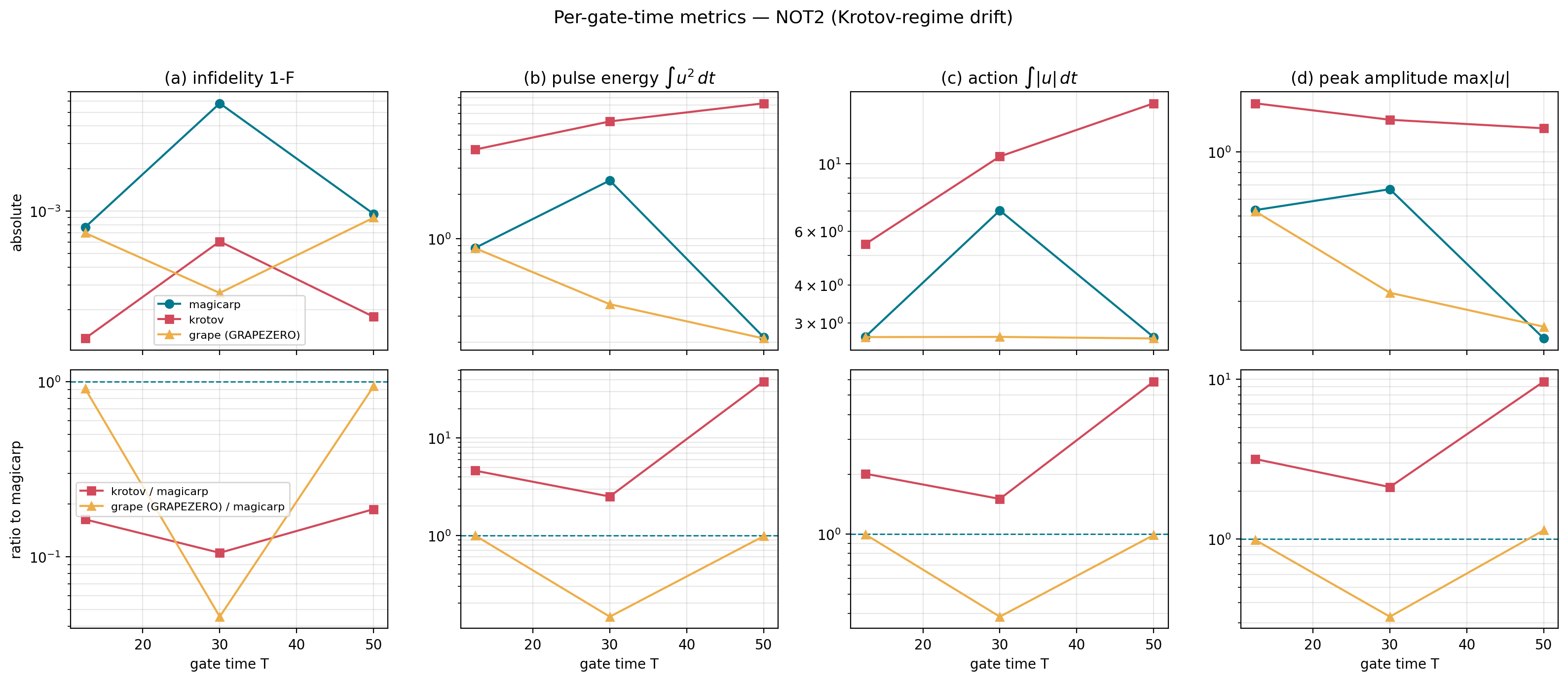}
\caption{Fair-halting comparison for $\mathrm{NOT}_2$ in the strong-coupling regime
($10^{-3}$ halt): verified infidelity, pulse energy $E_2$, and peak amplitude versus gate
time. GRAPE reproduces the MAGICARP energy curve; Krotov converges everywhere at an
order-of-magnitude energy premium.}
\label{fig:krotovreg-fair1e3-perT}
\end{figure}

\begin{figure}[htbp]
\centering\includegraphics[width=0.72\linewidth,trim=0 0 0 50,clip]{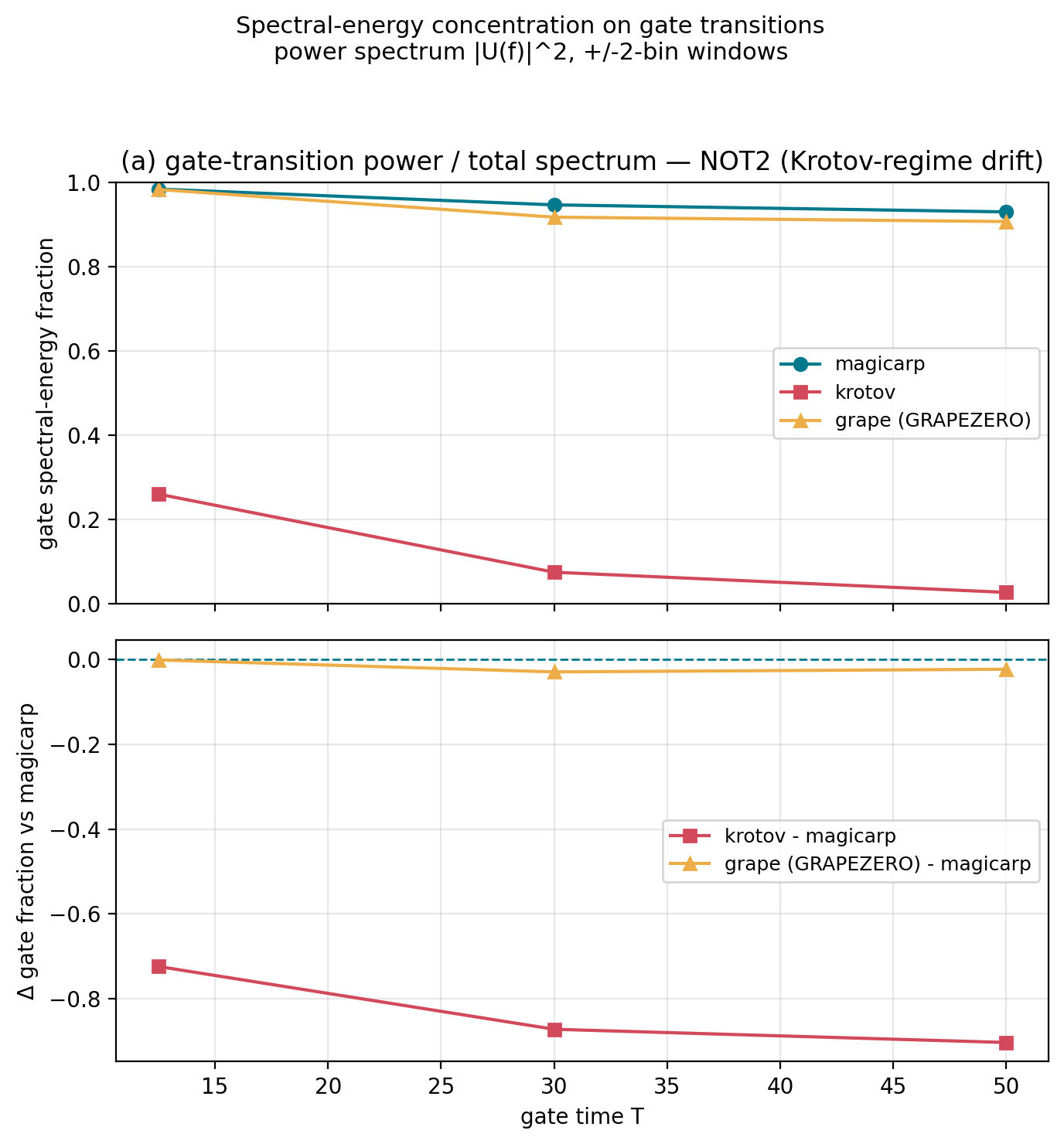}
\caption{Gate-tone spectral fraction for the three methods in the strong-coupling regime.
At the $10^{-3}$ halt the MAGICARP and GRAPE pulses are equally gate-concentrated
(spectator weight $\sim\!10^{-4}$); the Krotov pulse remains broadband at every gate
time.}
\label{fig:krotovreg-fair1e3-specfrac}
\end{figure}

\section{Single-Method Production Results}
\label{sec:app-solo}

For reference, this section collects the \emph{production} (best-of-search)
results of the drift-aware workflow alone --- without the fair-halting
constraint of \cref{sec:methods-fair} --- across the strong-coupling $\mathrm{NOT}_2$ benchmark and the moderate-exchange dressed QFT,
together with the grid-convergence and robustness studies that accompany them.

\subsection{Benchmark against Krotov: $\mathrm{NOT}_2$ on an exchange-coupled surface qubit}
\label{sec:krotov-compare}

To place the drift-aware MAGICARP construction on the same footing as a
state-of-the-art gradient method, we reproduce the
$\mathrm{NOT}_2$ benchmark study of the static exchange-coupled surface-qubit
platform of Ref.~\cite{surfaceqb2026}, performed there for closed-system dynamics, at the surface-qubit operating point of
Section~2.4 of the main text. Time evolution is computed by the same Trotterized
matrix exponential at $\Delta t = 0.01\,\mathrm{ns}$ used for the reference
Krotov pulses, so the two methods are compared under identical discretization.

\begin{table}[htbp]
\centering
\caption{Drift-aware MAGICARP $\mathrm{NOT}_2$ in the strong-coupling regime: converged
fidelity, integrated pulse power $E_2=\int_0^T|u_{\mathrm{lab}}|^2\,dt$, peak amplitude, and
the ratio of mean dressed-envelope RMS on the two gate transitions to that on the two
spectator transitions.}
\label{tab:krotov}
\begin{tabular}{ccccc}
\toprule
$T$ (ns) & $\mathcal{F}$ & $E_2$ (rad$^2$/ns) & $\|u_{\mathrm{lab}}\|_\infty$ (rad/ns) & gate/spectator \\
\midrule
50   & 0.99993 & 0.219 & 0.140 & $9.5\times$ \\
30   & 0.99988 & 0.362 & 0.222 & $47\times$ \\
12.5 & 0.99930 & 0.870 & 0.537 & $40\times$ \\
\bottomrule
\end{tabular}
\end{table}

\paragraph{Fidelity.} At the reference gate time $T=50\,\mathrm{ns}$ the shooting pulse reaches
$\mathcal{F}=0.99993$ (process fidelity; the corresponding average gate fidelity is
$\tfrac{d\mathcal{F}+1}{d+1}=0.99994$), matching the near-unit fidelity obtained by the Krotov
method in the same closed system~\cite{surfaceqb2026}. High fidelity is retained down to the
shortest experimentally relevant duration $T=12.5\,\mathrm{ns}$ ($\mathcal{F}=0.99930$). The
drift-aware MAGICARP pulse is therefore not a low-fidelity, smoothness-constrained alternative:
in this weak-driving regime ($\|u_{\mathrm{lab}}\|_\infty/\omega \sim 10^{-2}$) the
dressed-transition rotating-frame initialization is highly accurate, and the laboratory-frame
refinement closes the residual counter-rotating and cross-resonance errors.

\paragraph{Spectral parsimony versus Krotov.} The qualitative difference appears in the control
spectrum (not shown). The Krotov pulse, initialized from a broadband flat-top,
distributes spectral weight over \emph{all four} single-flip resonances $\omega_{\mathrm{RF}1\dots4}$.
The drift-aware MAGICARP pulse instead concentrates almost all of its weight on the
\emph{two} gate-essential dressed transitions and suppresses the spectators by one to nearly two
orders of magnitude (\cref{tab:krotov}). This selectivity is
structural rather than imposed: by Eq.~(10) of the main text the reconstructed pulse is, by
construction, a sum of carriers at the dressed transition frequencies modulated by slowly varying
envelopes, and the optimizer recruits the spectator tones only to the small extent needed to cancel
coherent cross-talk. For a comparable gate fidelity, the MAGICARP pulse thus addresses fewer
transitions and concentrates its energy near the gate-relevant resonances; the quantitative,
fair-halted version of this comparison is given in
\cref{sec:krotovreg-fair-1e3} and \cref{tab:krotovreg-fair}.

\paragraph{Control cost and robustness.} The pulse energy $E_2$ and peak amplitude both grow as
$\sim 1/T$ as the gate is compressed (not shown), the expected energy--time
trade-off~\cite{caneva2009}, while the integrated pulse area is essentially fixed by the $\pi$
rotation. Holding the optimized pulse fixed and perturbing the exchange coupling
shows that shorter gates are markedly more tolerant of a $J$ error,
since the always-on exchange has less time to act coherently during a fast gate; this
static-robustness gain runs opposite to the energy cost.

\paragraph{Time-grid convergence at $T=50\,\mathrm{ns}$.}
Re-optimizing on progressively finer grids at $T=50\,\mathrm{ns}$
shows the infidelity near its optimizer floor at $\Delta t=0.04\,\mathrm{ns}$ ($\sim1.8\times10^{-3}$),
converging to $\approx6.7\times10^{-5}$ by $\Delta t=0.02\,\mathrm{ns}$ with no further improvement
at $0.01\,\mathrm{ns}$. The dressed carriers at $6$--$10\,\mathrm{GHz}$ are adequately resolved
at $\Delta t=0.02\,\mathrm{ns}$, consistent with the Krotov-reference discretization of
$\Delta t=0.01\,\mathrm{ns}$.

\paragraph{Numerical note.} Reaching the parsimonious, near-unit-fidelity solution requires bounding
the shooting amplitude. With an overly loose bound the laboratory-frame refinement is attracted to a
high-amplitude basin ($\|u_{\mathrm{lab}}\|_\infty \sim 9\,\mathrm{rad/ns}$) whose spectrum is smeared
across all four resonances and whose fidelity saturates near $0.965$; constraining the generator
recovers the low-energy two-tone solution reported here (cf.\ the systematic mapping of this
failure mode in Section~4 of the main text). All fidelities are the verified
laboratory-frame values; energy and spectral quantities are evaluated a posteriori.


\subsection{Dressed QFT at moderate exchange}
\label{sec:realistic-qft}

We finally apply the identical drift-aware MAGICARP workflow, with the same
drift and scalar control, to the dressed two-qubit QFT of
Section~2.4 of the main text. In contrast to $\mathrm{NOT}_2$, which only flips two
dressed transitions, the QFT is a fully entangling gate that connects all four
levels.

\begin{table}[htbp]
\centering
\caption{Drift-aware MAGICARP dressed QFT at the moderate-exchange point. The gate is realized to
machine precision for $T\ge 25\,\mathrm{ns}$ but degrades at the shortest duration. Units as in
Table~2 of the main text.}
\label{tab:qft}
\begin{tabular}{ccccc}
\toprule
$T$ (ns) & $1-\mathcal{F}$ & $E_2$ & $E_1$ & $\|u_{\mathrm{lab}}\|_\infty$ \\
\midrule
50   & $5\times10^{-14}$ & 0.354 & 3.51 & 0.189 \\
25   & $3\times10^{-14}$ & 0.777 & 3.66 & 0.402 \\
12.5 & $3.1\times10^{-2}$ & 2.00 & 4.02 & 1.32  \\
\bottomrule
\end{tabular}
\end{table}

\paragraph{A gate-time threshold.} For $T=50\,\mathrm{ns}$ and $T=25\,\mathrm{ns}$ the dressed QFT is synthesized to
numerical machine precision ($1-\mathcal{F}\sim10^{-14}$) with smooth, low-amplitude pulses
($\|u_{\mathrm{lab}}\|_\infty \le 0.4\,\mathrm{rad/ns}$), demonstrating that the shooting parametrization extends
naturally from single-qubit-flip gates to fully entangling gates. At $T=12.5\,\mathrm{ns}$, however, the best
fidelity, obtained from a search over amplitude bounds and random restarts, reaches only
$\mathcal{F}\approx0.97$ ($1-\mathcal{F}\approx3\times10^{-2}$). This residual does not close when
the amplitude bound is varied (intermediate bounds even fall into a high-amplitude basin near
$\mathcal{F}\approx0.5$) nor when the integration grid is refined. The
limitation is therefore not numerical: it reflects a gate-time threshold for realizing the full QFT
with a single scalar control and this moderate exchange, consistent with general
quantum-speed-limit arguments~\cite{caneva2009,deffner2017}; the canonical treatment of this
threshold --- its sharp onset at $T\approx14\,\mathrm{ns}$, its identification with the addressing time of the single shared control, and the minimum-energy law above it --- is the sweep of
Section~4 of the main text.

\paragraph{The QFT uses all dressed transitions.} The dressed-transition usage
and the control spectrum (not shown; cf.\ Figure~4 of the main text) show that, unlike
the parsimonious $\mathrm{NOT}_2$ pulse, the QFT drives all four dressed transitions with comparable
amplitude and produces a four-peak spectrum. This is the expected counterpart of the spectral
parsimony observed for $\mathrm{NOT}_2$: the ``useful set'' of transitions for the QFT is the entire
dressed-transition manifold, so the optimizer correctly populates every resonance. The
energy--time trade-off is preserved: $E_2$ and the peak amplitude grow
as the gate is compressed, while the pulse area $E_1\approx3.5\,\mathrm{rad}$ is roughly conserved and is larger
than for $\mathrm{NOT}_2$ ($\approx2.7\,\mathrm{rad}$), consistent with the QFT being a ``larger'' rotation.
At $T=50\,\mathrm{ns}$ the grid convergence is dramatic: the infidelity plunges from
$\sim3\times10^{-2}$ at $\Delta t=0.1\,\mathrm{ns}$ to $\sim3\times10^{-8}$ at $\Delta t=0.05\,\mathrm{ns}$ and machine
precision by $\Delta t=0.02\,\mathrm{ns}$, reflecting the threshold character of the QFT synthesis --- once the
grid resolves the dressed carriers sufficiently, the optimizer reaches exact synthesis.

\paragraph{Robustness.} As for the other gates, the fixed-pulse fidelity at $T=50\,\mathrm{ns}$ is sensitive to a
uniform error in the exchange coupling; the resonant, four-tone pulse is
tuned to the dressed transition frequencies and degrades when a change in $J$ detunes them.


\end{document}